\documentclass[twoside,twocolumn]{article}
\usepackage[super,sort&compress,comma]{natbib} 
\usepackage{amsmath,amssymb}
\usepackage{times,mathptmx}
\usepackage{sectsty}
\usepackage{graphicx}
\usepackage{fancyhdr}
\begin{document}

\renewcommand{\thefootnote}{\fnsymbol{footnote}}
\renewcommand\footnoterule{\vspace*{1pt}%
\hrule width 3.4in height 0.4pt \vspace*{5pt}} 
\setcounter{secnumdepth}{5}

\makeatletter 
\def\subsubsection{\@startsection{subsubsection}{3}{10pt}{-1.25ex plus -1ex minus -.1ex}{0ex plus 0ex}{\normalsize\bf}} 
\def\paragraph{\@startsection{paragraph}{4}{10pt}{-1.25ex plus -1ex minus -.1ex}{0ex plus 0ex}{\normalsize\textit}} 
\renewcommand\@biblabel[1]{#1}            
\renewcommand\@makefntext[1]%
{\noindent\makebox[0pt][r]{\@thefnmark\,}#1}
\makeatother 
\renewcommand{\figurename}{\small{Fig.}~}
\sectionfont{\large}
\subsectionfont{\normalsize} 

\renewcommand{\headrulewidth}{1pt} 
\renewcommand{\footrulewidth}{1pt}
\setlength{\arrayrulewidth}{1pt}
\setlength{\columnsep}{6.5mm}
\setlength\bibsep{1pt}

\twocolumn[
  \begin{@twocolumnfalse}
\noindent\LARGE{\textbf{Mathematical diffraction of aperiodic structures$^\dag$}}
\vspace{0.6cm}

\noindent\large{\textbf{Michael Baake\textit{$^{a}$} and
Uwe Grimm\textit{$^{b}$}}}\vspace{0.5cm}

\noindent \normalsize{Kinematic diffraction is well suited for a
  mathematical approach via measures, which has substantially been
  developed since the discovery of quasicrystals. The need for further insight
  emerged from the question of which distributions of matter, beyond
  perfect crystals, lead to pure point diffraction, hence to
  sharp Bragg peaks only.  More recently, it has become apparent
  that one also has to study continuous diffraction in more
  detail, with a careful analysis of the different types of diffuse
  scattering involved.  In this review, we summarise some key results,
  with particular emphasis on non-periodic structures. We choose an
  exposition on the basis of characteristic examples, while we refer
  to the existing literature for proofs and further details.}
\vspace{0.5cm}
 \end{@twocolumnfalse}
  ]

\footnotetext{\dag~Part of a themed issue on Quasicrystals in honour of the 2011 
Nobel Prize in Chemistry winner, Professor Dan Shechtman.}
\footnotetext{\textit{$^{a}$~Fakult\"{a}t f\"{u}r Mathematik, 
       Universit\"{a}t Bielefeld,
       Postfach 100131, 33501 Bielefeld, Germany. 
       E-mail: mbaake@math.uni-bielefeld.de}}
\footnotetext{\textit{$^{b}$~Department of Mathematics and Statistics, 
       The Open University,
       Walton Hall, Milton Keynes MK7 6AA, United Kingdom. 
       Email: u.g.grimm@open.ac.uk}}

\section{Introduction}

Diffraction techniques have dominated the structure analysis of solids
for the last century, ever since von Laue and Bragg employed X-ray
diffraction to determine the atomic structure of crystalline
materials. Despite the availability of direct imaging techniques such
as electron and atomic force microscopy, diffraction by X-rays,
electrons and neutrons continues to be the method of choice to detect
order in the atomic arrangements of a substance; see Cowley's
book\cite{C} and references therein for background.

In its full generality, the diffraction of a beam of X-rays, electrons
or neutrons from a macroscopic piece of solid is a complicated
physical process.  It is the presence of inelastic and multiple
scattering, prevalent particularly in electron diffraction, which
makes it essentially impossible to arrive at a complete mathematical
description of the process. Here, we restrict to kinematic diffraction
in the far-field or Fraunhofer limit. In this case, powerful tools of
harmonic analysis are available to attack the direct problem of
calculating the (kinematic) diffraction pattern of a given structure. 

In contrast, the \emph{inverse problem} of determining a structure
from its diffraction intensities is extremely involved. A diffraction
pattern rarely determines a structure uniquely, as there can be
\emph{homometric} structures sharing the same autocorrelation (and
hence the same diffraction).\cite{Pat44,GM95,BG07,GB08} We are far
away from a complete understanding of the homometry classes of
structures, in particular if the diffraction spectrum contains
continuous components. At present, a picture is emerging, based on the
analysis of explicit examples, which highlight how large the homometry
classes may be.

\begin{figure}[t]
\centering
\includegraphics[width=0.9\columnwidth]{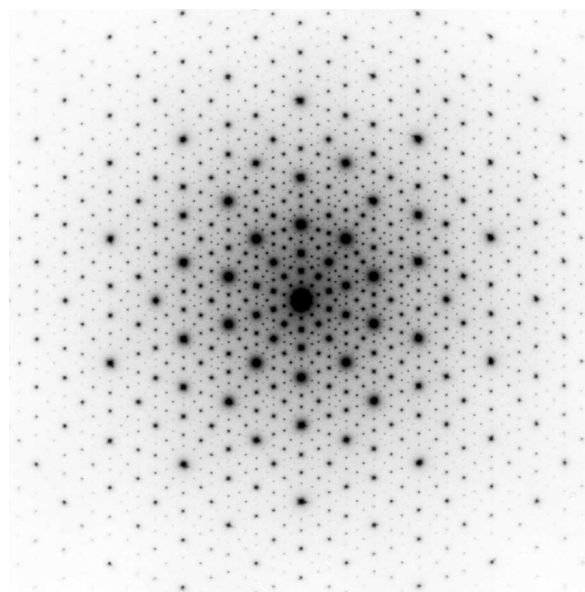}
  \caption{~Experimental diffraction pattern of a quasicrystalline 
  AlPdMn alloy.
  Figure courtesy of Conradin Beeli.}
  \label{fig:icoAMP}
\end{figure}

Originally, much of the effort concentrated on the pure point part of
diffraction, also called the Bragg diffraction, for the case of
ordinary (periodic) crystals, and later also for incommensurate
phases.  Following the discovery of
quasicrystals\cite{SBGC84,INF85,KN,LevSt} with their beautiful
diffraction patterns, such as the one shown in
Figure~\ref{fig:icoAMP}, a new mathematical approach was required. The
associated paradigm shift also re-opened the discussion of what
possible manifestations of order and disorder in solids there are, and
how these can be detected and quantified. While diffraction is one
measure of order, the existence of homometric structures of varying
entropy\cite{BG09,BG12} shows its limitations, as there are completely
deterministic systems which cannot be distinguished from a randomly
disordered system on the basis of pair correlations
alone. Increasingly, the continuous or diffuse part of the diffraction
is attracting attention,\cite{WW,Withers,MdB} not the least because
improved experimental techniques make the diffuse part
accessible. Improving our understanding of diffuse diffraction is
desirable, in particular in view of the implications on disorder.

\begin{figure}[t]
\centering
\includegraphics[width=\columnwidth]{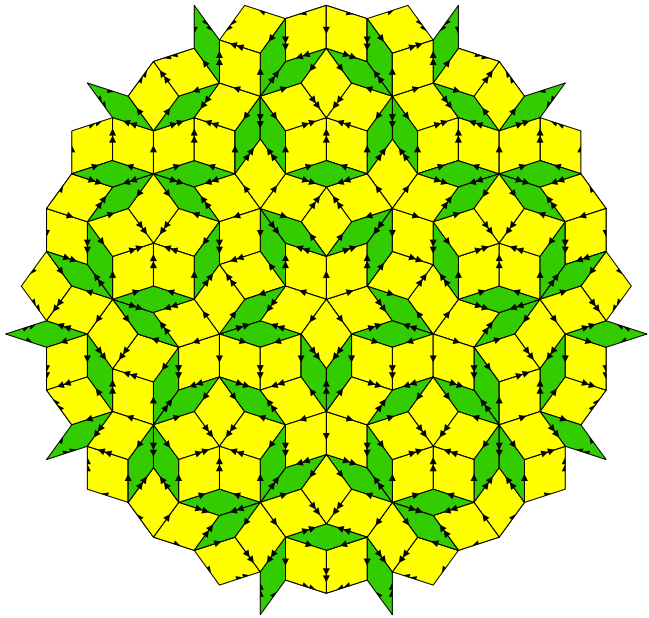}
  \caption{~A patch of the rhombic Penrose tiling. The arrow decorations 
of the edges encode the local rules.}
  \label{fig:pen}
\end{figure}

The most successful approach to describe the structure of
incommensurate crystals and quasicrystals employs additional
dimensions. By embedding the ideal structure into a higher-dimensional
`superspace', it is possible to recover periodicity in the
higher-dimensional space, and this picture can be extended to cover
certain aspects of random tilings as well.  The standard tilings used
to model the structure of quasicrystals are obtained in this way; for
instance, the Penrose tiling\cite{Pen74} shown in Figure~\ref{fig:pen}
can be described as a projection\cite{BKSZ90} of a slice through the
four-dimensional root lattice $A_{4}$. Such structures, or their
equivalent point sets, are called \emph{cut and project sets} or
\emph{model sets}, and we shall discuss further examples below. Note
that the Penrose tiling also possesses aperiodic, perfect \emph{local
  rules} (or matching rules\cite{Katz,Soc90,Goo98}), as well as an
inflation symmetry. The local rules can be implemented as arrow
decorations on the edges of the two rhombic prototiles, which, within
any admissible patch, have to agree on all edges. These local rules
are aperiodic in the sense that they are incompatible with any
periodic tiling. They are perfect because they specify precisely the
class of the rhombic Penrose tilings, in the sense that all
space-filling tilings obeying these rules are locally
indistinguishable (LI) from the rhombic Penrose tiling, the latter
defined as a fixed point tiling of an inflation rule.

\begin{figure}[t]
\centering
\includegraphics[width=\columnwidth]{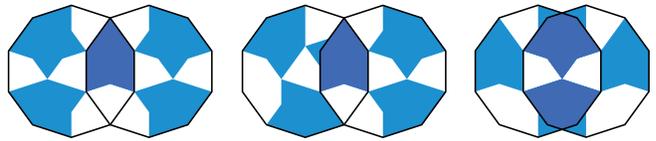}
\caption{~The three allowed (pairwise) overlaps of the decagonal
  cluster. Overlapping markings are highlighted by colour.}
  \label{fig:overlap}
\end{figure}

\begin{figure}[t]
\centering
\includegraphics[width=\columnwidth]{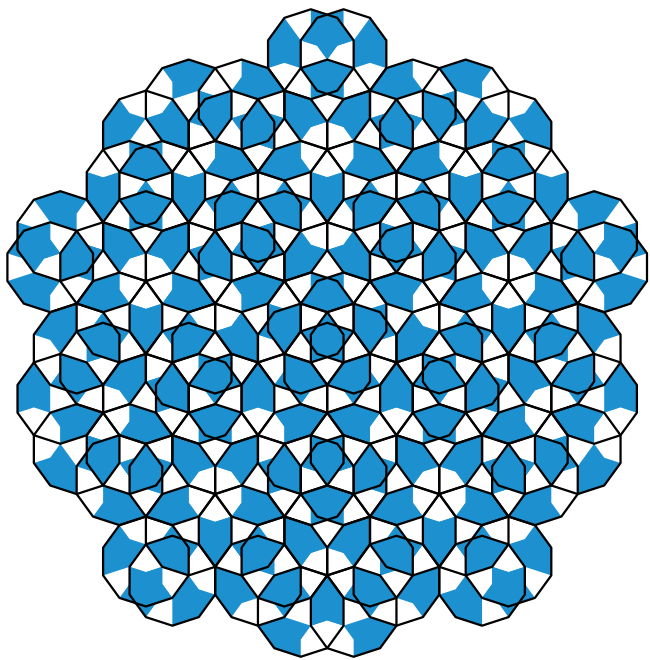}
  \caption{~A patch of Gummelt's decagon covering.}
  \label{fig:cover}
\end{figure}

It is worth noting that, while the lattice of periods of a periodic
crystal is unique (though the choice of unit cell is not), there is
considerable freedom in the choice of the building blocks of aperiodic
tilings. In the case of the Penrose tiling, there exist a number of
equivalent versions (in the sense\cite{BSJ,B,TAO} of mutual local
derivability), such as the Penrose pentagon tiling or the kite and
dart tiling. One can even go beyond tilings and consider coverings of
space.\cite{KPBook} In the case of the Penrose tiling, Gummelt's
decagon covering\cite{Gum96} with a single cluster (and overlap rules
encoded by the shading) has proved very popular, because it allows the
description of a quasicrystal structure in terms of a single
fundamental building block. The three allowed (pairwise) overlaps of
the marked decagons, shown in Figure~\ref{fig:overlap}, are
characterised by matching decorations. Figure~\ref{fig:cover} shows a
patch of a corresponding covering, which is mutually locally derivable
(MLD) with the Penrose tiling of
Figure~\ref{fig:pen}.\cite{Gum96,GumBook} This covering also has an
interpretation in terms of `maxing rules',\cite{JS94,Hen97,GGB03}
where maximisation of one type of specified cluster leads to the
Penrose rhombus tiling (up to zero density deviations).\cite{JS94}
Covering rules of either type have become quite fashionable in
materials science.\cite{RG03,Steu04} For more examples on tilings, in
particular on substitution tilings, we refer to the online Tilings
Encyclopedia.\cite{encyc} For the early development of the field, the
reprint volume by Steinhardt and Ostlund\cite{SO} is still a valuable
source.

This review attempts to present an overview of the development of
mathematical diffraction theory in the 30 years since the discovery of
quasicrystals by Shechtman et al.\cite{SBGC84} While we aim to provide
the reader with a flavour of the mathematical methods and assumptions,
we will not dive deeply into the technical details. In particular, we
will not present any formal proofs, though we do state several
non-trivial results explicitly. We refer to our recent
review\cite{diffrev} and our forthcoming book,\cite{TAO} and the
references contained therein, for more details on the rigorous
mathematical treatment. Three complementary review
volumes\cite{Nato,BMBook,Beyond} with mathematical articles are also
highly recommended. Here, we select examples that are both
characteristic and somewhat supplementary to previous presentations.

In Section~\ref{sec:methods}, we start with a concise summary of the
systematic approach using \emph{measures} (in the mathematical sense,
such as Lebesgue measure $\lambda$, which is used to measure volume in
Euclidean space), which was pioneered in this context by
Hof.\cite{Hof,Hof2,Hof97} We first apply this approach to the
diffraction of perfect crystals in Section~\ref{sec:crystals}, and
then discuss the case of mathematical quasicrystals based on a cut and
project scheme in Section~\ref{sec:qc}.  Like perfect (or idealised)
crystals, these systems are pure point diffractive, which means that
the diffraction pattern consists of sharp (Bragg) peaks only.
Afterwards, in Section~\ref{sec:cont}, we proceed to systems with
continuous diffraction, covering both the case of singular continuous
and absolutely continuous diffraction by means of representative
examples, including a probabilistic model for thermal fluctuations. In
particular, we consider random tilings, which are relevant because
most quasicrystalline materials show entropic stabilisation and
therefore are expected to include configurational disorder.

\section{Methods and general results}
\label{sec:methods}

For a satisfying mathematical approach, we should exclude any boundary
effects, and hence consider infinite systems that represent the
scattering medium.  Traditionally, there are two seemingly
contradictory ways to describe a system, either in terms of
\emph{functions} which represent the density of the scattering medium,
or by lattices or, more generally, \emph{tilings} of space, whose
decorations mimic the atomic positions. This dichotomy has sparked
some rather fierce disputes between the tiling school and the density
function school, in particular in the years following the discovery of
quasicrystals. However, the two viewpoints can be reconciled by
embedding them into a more general frame. One way of doing that is to
introduce \emph{measures}, which comprise (almost) periodic functions
and tilings as special cases. As measures quantify distributions in
spaces, this approach is in fact very natural, and well suited to
describe both the distribution of matter in the scattering medium and
the distribution of (scattered) intensity in space. We therefore start
by briefly introducing the concepts and main properties that will be
needed in our context.

\subsection{Measures, convolutions and Fourier transforms}

Due to the Riesz-Markov representation theorem,\cite{RS} it is
possible to think of a measure as a linear functional, i.e., as a
linear map that associates a number to each function from an
appropriate space. A (complex) measure $\mu$ on $\mathbb{R}^{d}$ is
then a linear functional (with values in the complex numbers
$\mathbb{C}$) on the space $C_{\mathrm{c}}(\mathbb{R}^{d})$ of
complex-valued, continuous (test) functions of compact support,
subject to the condition that, for every compact set
$K\subset\mathbb{R}^{d}$, there is a constant $a^{}_{K}$ such that
\[
    \lvert \mu(g)\rvert \, \le \, a^{}_{K}\, \lVert g\rVert^{}_{\infty}
\]
for all test functions $g$ with support in $K$. Here, $\lVert
g\rVert^{}_{\infty}=\sup_{x\in K}\lvert g(x)\rvert$ is the supremum
norm of $g$. 

We write $\mu(g)$ or $\int_{\mathbb{R}^{d}} g(x)\, \mathrm{d}\mu(x)$
for the measure of a function $g$, and $\mu(A)=\mu(1^{}_{A})$ for the
measure of a set $A\subset\mathbb{R}^{d}$, where
\[  
   1^{}_{A}(x)\, =\, \begin{cases} 1, & \mbox{if $x\in A$,}
\\ 0, & \mbox{otherwise,} \end{cases}
\]
denotes the \emph{characteristic function} of the set $A$.

If $\mu$ is a complex measure, the \emph{conjugate} of $\mu$ is the
measure $\bar{\mu}$ which is defined by
$g\mapsto\overline{\mu(\bar{g})}$.  A measure is called \emph{real}
(or signed), when $\bar{\mu}=\mu$, and it is called \emph{positive}
when $\mu(g)\ge 0$ for all $g\ge 0$. For every measure $\mu$, there is
a smallest positive measure, denoted by $\lvert \mu\rvert$, such that
$\lvert\mu(g)\rvert\le\lvert\mu\rvert(g)$ for all non-negative
$g$. This is called the \emph{total variation} (or absolute value) of
$\mu$. A measure $\mu$ is called finite or \emph{bounded}, if
$\lvert\mu\rvert(1)=\lvert\mu\rvert(\mathbb{R}^{d})$ is finite,
otherwise it is called unbounded. As we want to describe infinite
point sets in space, we usually deal with the latter case, but we will
assume that measures are \emph{translation bounded}.  This means that,
for any compact set $K\subset\mathbb{R}^{d}$, the total variation
satisfies
\[
   \sup_{t\in\mathbb{R}^{d}} \,\lvert \mu\rvert (t+K) \; 
    < \; \infty\, ,
\]
so wherever you move your compact set $K$, its total variation measure
is always finite. 

\subsection{Autocorrelation and diffraction measures}

If $\varLambda\subset\mathbb{R}^{d}$ is a point set that is a Delone
set (a set where points neither get arbitrarily close nor so sparse
that it accommodates arbitrarily large empty balls), the corresponding
\emph{Dirac comb}\cite{Cor}
\[
    \delta_{\varLambda}\, := \, \sum_{x\in\varLambda} \delta_{x}
\]
is a translation bounded measure, where $\delta_{x}$ is the normalised
(Dirac) point measure at $x$ (so $\delta_{x}(g)=g(x)$, or, in the
formal notation used in physics, $\int_{\mathbb{R}^{d}} g(y)\,
\delta (y-x)\, \mathrm{d}y = g(x)$). In what follows, we use such Dirac
combs to represent the scattering medium, possibly with (in general
complex) scattering weights $w(x)$ at position
$x\in\mathbb{R}^{d}$. The corresponding weighted Dirac comb is denoted
as
\[
    \omega \, = \, 
    w\, \delta_{\varLambda}\, = \, 
    \sum_{x\in\varLambda} w(x)\, \delta_{x}\, .
\]

If $\omega$ is a translation bounded measure, the corresponding
diffraction measure is the Fourier transform of the autocorrelation
measure, where we shall assume that the latter exists. In any given
example, this has to be verified, of course.  The
\emph{autocorrelation measure} of $\omega$ is defined as the limit
\begin{equation}\label{eq:def-gamma}
   \gamma \, =  \, \gamma^{}_{\omega} 
                  = \, \omega \circledast \widetilde{\omega} 
                 \, := \lim_{R\to\infty}
                \frac{\;\omega|^{}_{R} \ast \widetilde{\omega|^{}_{R}}\;}
                   {\mathrm{vol} (B_R)} ,
\end{equation}
where $B_{R}$ denotes the open ball of radius $R$ around $0\in
\mathbb{R}^{d}$. By $\omega|^{}_{R}$ we denote the restriction of
$\omega$ to the ball $B_{R}$. For a measure $\mu$, its `flipped-over'
version $\widetilde{\mu}$ is defined via $\widetilde{\mu}(g) =
\overline{\mu(\widetilde{g})}$, where
$\widetilde{g}(x)=\overline{g(-x)}$.  The operation $\ast$ is the
ordinary \emph{convolution} of measures, which is a generalisation
of the standard convolution of integrable functions,
\[
    \bigl(f\ast g\bigr)(x)\,  := 
    \int_{\mathbb{R}^d} f(x-y)\, g(y) \, \mathrm{d}y
    \, = \int_{\mathbb{R}^d} f(y)\, g(x-y) \, \mathrm{d}y \, .
\]
For finite measures $\mu$ and $\nu$ on $\mathbb{R}^{d}$, it is defined
by
\[
   \bigl(\mu\ast\nu\bigr)(g)\, = \int_{\mathbb{R}^{d}\times\mathbb{R}^{d}}
   g(x+y) \, \mathrm{d}\mu (x) \, \mathrm{d}\nu (y) 
\]
for any function $g\in C_{\mathrm{c}}(\mathbb{R}^{d})$, which is then
again a finite measure.  The volume-averaged convolution $\circledast$
(also called the \emph{Eberlein convolution}, in analogy to a similar
approach\cite{GLA} in the theory of almost periodic measures) is
needed in Eq.~\eqref{eq:def-gamma}, because $\omega$ itself is
generally an unbounded measure and the direct convolution is not
defined. For example, if $\lambda$ denotes the standard Lebesgue
measure (for volume), $\lambda\ast\lambda$ is not defined, while
$\lambda\circledast\lambda=\lambda$.

\begin{figure}[t]
\centering
\includegraphics[width=0.9\columnwidth]{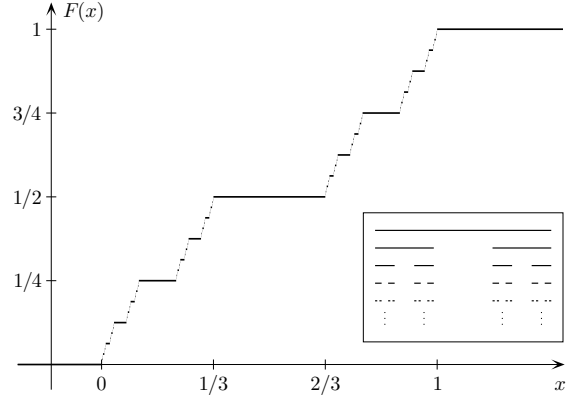}
  \caption{~Illustration of the distribution function $F(x)$
of the classic middle-thirds Cantor set. The iterative construction
for the latter is sketched in the inset.}
  \label{fig:cantor}
\end{figure}

If the autocorrelation measure $\gamma$ of $\omega$ exists, its
Fourier transform $\widehat{\gamma}$ does as well, and
$\widehat{\gamma}$ is a translation bounded, positive measure, called
the \emph{diffraction measure} of $\omega$. It corresponds to the 
kinematic scattering intensity observed in an experiment in the sense
that it quantifies how much scattering intensity reaches a given
volume in $d$-space. Relative to Lebesgue measure $\lambda$, the
diffraction measure has a unique decomposition\cite{RS}
\[
   \widehat{\gamma} \; = \; 
            \widehat{\gamma}^{}_{\mathrm{pp} } +
            \widehat{\gamma}^{}_{\mathrm{sc} } +
            \widehat{\gamma}^{}_{\mathrm{ac} }
\]
into its pure point part (the Bragg peaks, of which there are at most
countably many), its absolutely continuous part (the diffuse
background scattering, which has a locally integrable density relative
to $\lambda$) and its singular continuous part (which simply means
anything that remains, which is nothing in many standard cases
considered in crystallography). Each of the three terms is again a
positive measure.  Singular continuous measures are weird objects:\
they give no weight to single points, but are still concentrated to an
(uncountable!) set of zero Lebesgue measure. A well-known example is
the probability measure for the classic middle-thirds Cantor
set,\cite{RS} with the 'Devil's stair case' as its distribution
function, which is constant almost everywhere; see
Figure~\ref{fig:cantor}. Singular continuous diffraction does occur in
realistic models though,\cite{HB00} and should not be disregarded.

\section{Diffraction of perfect crystals}
\label{sec:crystals}

In our setting, a perfect (infinite) crystal in $d$-space is a
lattice-periodic (discrete) structure.  It is defined by its lattice
of periods $\varGamma\subset\mathbb{R}^{d}$ and the decoration of a
fundamental domain of $\varGamma$, which together completely specify
the distribution of scatterers in space. It is therefore described by
a crystallographic measure
\begin{equation}\label{eq:cryst}
    \omega\, =\, \mu \ast \delta^{}_{\varGamma}\, ,
\end{equation}
where $\mu$ is a finite measure. The latter can be chosen as the
restriction of $\omega$ to a fundamental domain of
$\varGamma$. Depending on the nature of $\mu$, the resulting measure
$\omega$ can be pure point or continuous (for instance, if $\mu$ is
the constant measure on the fundamental domain, $\omega$ would be
proportional to Lebesgue measure), or a mixture of both types. One can
think of the Dirac comb $\delta^{}_{\varGamma}$ as implementing the
lattice periodicity, while $\mu$ describes the distribution of
scatterers in a fundamental domain of $\varGamma$.

The autocorrelation of the crystallographic measure $\omega$ of
Eq.~\eqref{eq:cryst} is given by
\begin{equation}\label{eq:crystauto}
     \gamma = \mathrm{dens} (\varGamma)\, 
      (\mu \ast \widetilde{\mu}) \ast \delta^{}_{\varGamma}\, ,
\end{equation}
which follows by using the relation $\widetilde{\delta^{}_{\varGamma}}
=\delta^{}_{\varGamma}$ together with
$\delta^{}_{\varGamma}\circledast\delta^{}_{\varGamma}= \mathrm{dens}
(\varGamma)\,\delta^{}_{\varGamma}$. Here, $\mathrm{dens}(\varGamma)$
denotes the density (per unit volume) of the lattice $\varGamma$,
which is the reciprocal of the volume of its fundamental
domain. Consequently, $\gamma$ is also a $\varGamma$-periodic
measure. In order to obtain the corresponding diffraction measure, we
need to know how to calculate the Fourier transform of
lattice-periodic measures.

\subsection{Poisson's summation formula}

A powerful tool for the Fourier analysis of lattice-periodic measures
is the \emph{Poisson summation formula} (PSF). For a lattice
$\varGamma\subset\mathbb{R}^{d}$ (which means that $\varGamma$ is a
discrete subgroup of $\mathbb{R}^{d}$ such that the factor group
$\mathbb{R}^{d}/\varGamma$ is compact), the Fourier transform of the
corresponding Dirac comb $\delta_{\varGamma}$ is
\begin{equation}\label{eq:psf}
 \widehat{\delta^{}_{\varGamma}}
     \, = \, \mathrm{dens} (\varGamma) \, 
     \delta^{}_{\varGamma^{*}}\, ,
\end{equation}
where $\varGamma^{*}$ denotes the \emph{dual} or
\emph{reciprocal lattice} of $\varGamma$. The latter is defined by
\[
    \varGamma^{*} \, =\, \{ x\in\mathbb{R}^{d} \mid \mbox{$\langle x|
      y\rangle \in \mathbb{Z}$ for all $y \in \varGamma$} \} \, .
\]
Here and below, $\langle x|y\rangle$ denotes the scalar product of
$x,y\in\mathbb{R}^{d}$. Note that sometimes a factor $2\pi$ is
included in the definition of the reciprocal lattice, which we prefer
to incorporate in our definition of the Fourier transform. For a
suitable function $\phi$, our convention for Fourier transform is
\[
    \widehat{\phi} (k) \, :=  \int_{\mathbb{R}^d} 
    e^{-2\pi i \langle k| x\rangle} \, \phi (x)\, \mathrm{d}x ,
\]
where $k,x\in\mathbb{R}^{d}$ and again $\langle k|x\rangle$ denotes
their scalar product.  The Fourier transform $\widehat{\gamma}$ of a
positive definite measure $\gamma$ (which means that $\gamma
(g\ast\widetilde{g})\ge 0$ holds for all $g\in C_{\mathrm{c}}
(\mathbb{R}^{d})$) is defined as the unique extension\cite{RS,BF} of
the Fourier transform of functions. It is conveniently defined in 
the setting of tempered distributions,\cite{RS} which provide
concrete means to calculate the transforms.

By the Bochner-Schwartz theorem,\cite{RS} the diffraction measure is
then a translation bounded positive measure. In addition, we will make
use of the \emph{convolution theorem} for measures. This states that
if $\mu$ is a finite measure and $\nu$ a translation bounded measure
on $\mathbb{R}^{d}$, the convolution $\mu\ast\nu$ exists and is a
translation bounded measure.\cite{BF} If $\widehat{\nu}$ is not only a
tempered distribution, but itself also a measure, one has the
convolution identity $\widehat{\mu\ast\nu\,} =
\widehat{\mu}\,\widehat{\nu}$. The latter is then again a measure,
which is absolutely continuous relative to $\widehat{\nu}$, because
$\widehat{\mu}$ is a bounded, uniformly continuous function on
$\mathbb{R}^{d}$ in this case.

\subsection{Diffraction of crystallographic structures}

Using the PSF together with the convolution theorem, the Fourier
transform of the crystallographic autocorrelation measure $\gamma$ of
Eq.~\eqref{eq:crystauto} can be calculated as
\begin{equation}\label{eq:crystdiff}
     \widehat{\gamma} = \bigl(  \mathrm{dens} (\varGamma) \bigr)^{2}
     \, \big| \widehat{\mu} \big|^{2} \, \delta^{}_{\varGamma^{*}} \, .
\end{equation}
Clearly, this is a pure point measure, concentrated on the dual
lattice $\varGamma^{*}$. Note that $\big| \widehat{\mu} \big|^{2}$ is
a uniformly continuous and bounded function that is evaluated only at
points of the dual lattice $\varGamma^{*}$. While different admissible
choices for the measure $\mu$ (describing the same system) lead to
different such functions, they agree on all points of $\varGamma^{*}$,
so that the result does not depend on this choice. If $\widehat{\gamma}
(\{ k \}) = 0$ for some $k\in\varGamma^{*}$, one calls this an
\emph{extinction}. Extinctions are characteristic features of
further symmetries, also of generalised type.

\subsection{Planar $\sigma$-phases}

\begin{figure}[t]
\centering
\includegraphics[width=0.76\columnwidth]{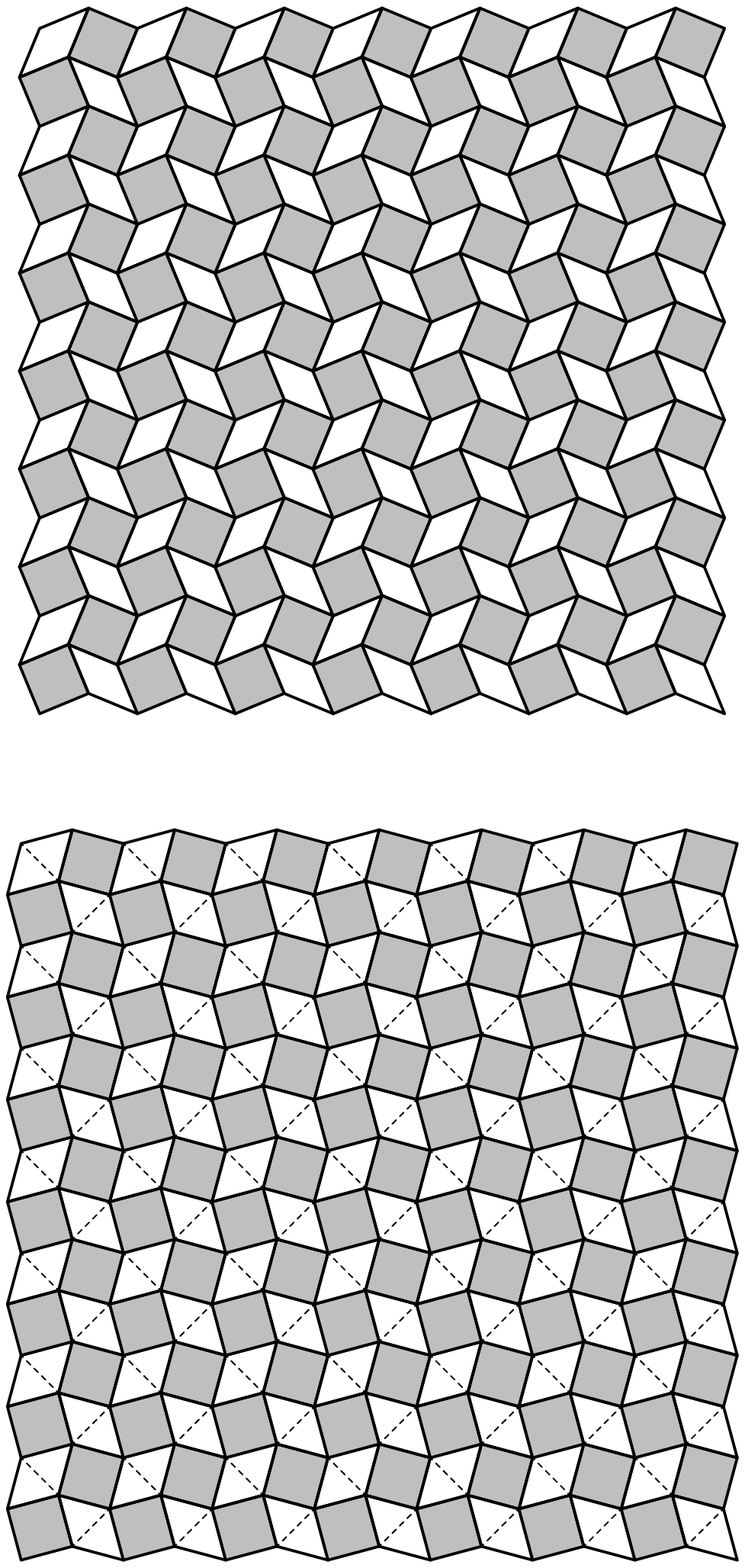}
  \caption{~Planar $\sigma$-phases with angles $\varphi=\pi/8$ (top) and
   $\varphi=\pi/12$ (bottom), shown with the correct relative length scale.
   In the latter case, the rhombus dissects into two equilateral triangles.}
  \label{fig:sigma}
\end{figure}

Let us consider an interesting example in some detail. Starting from a
checker board, viewed as a decoration of the square lattice, we assume
that the grey squares are stiff (or solid), while the white squares
are empty. One can now twist the structure by rotating the grey
squares alternately in opposite directions by an angle $\varphi\in
\bigl(-\frac{\pi}{4},\frac{\pi}{4}\bigr)$,
\[
\includegraphics[width=0.7\columnwidth]{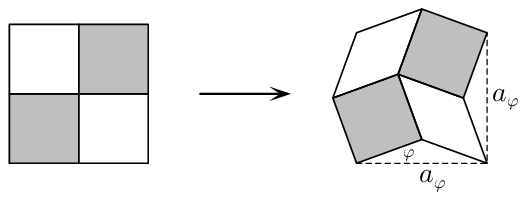}
\]
This way, a new periodic structure emerges where the white squares are
deformed into congruent rhombuses. This structure is the
lattice-periodic repetition of the motif above, and resembles a planar
$\sigma$-phase and related quasicrystal approximants.\cite{INF85} A
couple of examples are shown in Figure~\ref{fig:sigma}. The second
is related to structures found in $12$-fold symmetric 
quasicrystals.\cite{INF85}

We consider the associated Dirac comb 
\[
    \omega^{}_{\varphi} \, = \, \delta^{}_{R^{}_{\varphi} S} \,\ast\, 
    \delta^{}_{\alpha^{}_{\varphi} \mathbb{Z}^{2}}\, .
\]
obtained by placing a normalised point (or Dirac) measure at each
vertex point. Here, we have $\alpha^{}_{\varphi} = 2\cos (\varphi)$ and
$R^{}_{\varphi} = \left( \begin{smallmatrix} \cos (\varphi) & -\sin
  (\varphi) \\ \sin (\varphi) & \cos (\varphi) \end{smallmatrix}
\right)$, while $S = \{ 0, e^{}_{1}, e^{}_{2}, e^{}_{1} + e^{}_{2}\}$
denotes the vertex set of the unit square $[0,1]^{2}$. The corresponding
diffraction measure is obtained via Eq.~\eqref{eq:crystdiff} as
\[
   \widehat{\gamma^{}_{\varphi}} \, = \,
   \frac{1 + \cos \bigl(2 \pi 
        \langle R^{}_{\varphi} e^{}_{1} | k \rangle \bigr)}
        {2 \cos (\varphi)^{2}}\,
   \frac{1 + \cos \bigl(2 \pi 
        \langle R^{}_{\varphi} e^{}_{2} | k \rangle \bigr)}
        {2 \cos (\varphi)^{2}}\;
   \delta^{}_{\mathbb{Z}^{2}\!/2 \cos (\varphi)} ,
\]
with $\langle x | y \rangle$ denoting the scalar
product in $\mathbb{R}^{2}$.

\begin{figure}[t]
\centering
\includegraphics[width=0.76\columnwidth]{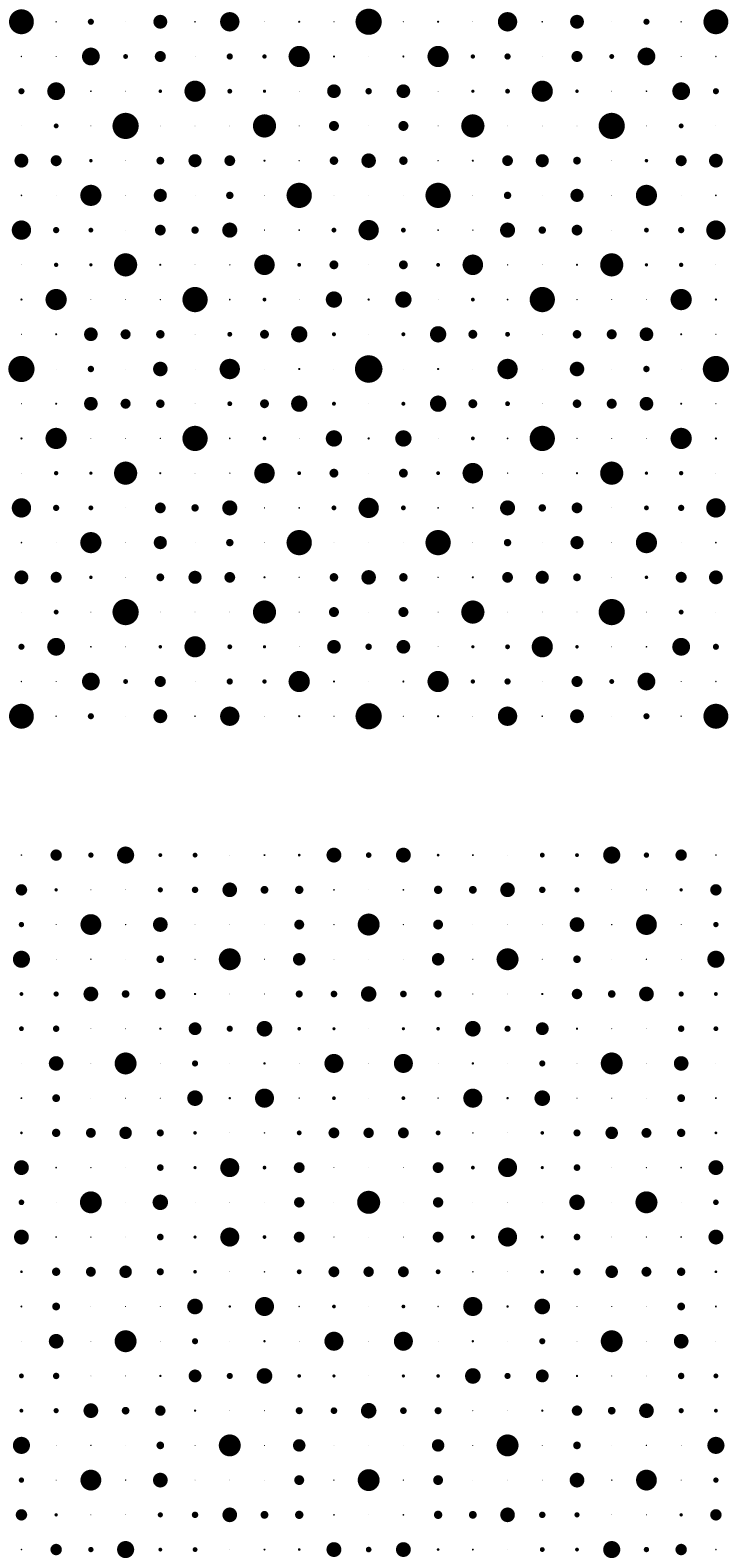}
  \caption{~Diffraction patterns for the two $\sigma$-phases of 
  Figure~\ref{fig:sigma}. All distances and intensities are shown
  in the correct relative scale.}
  \label{fig:sigmadiff}
\end{figure}

When $\varphi = 0$ (which means we are back to the square lattice),
this expression reduces to $\widehat{\gamma^{}_{0}} =
\delta^{}_{\mathbb{Z}^{2}}$, as it must, while inserting $\varphi =
\pm\pi/4$ leads to $\widehat{\gamma^{}_{\pm\pi/4}} = 4
\delta^{}_{R^{}_{\pi/4} \mathbb{Z}^{2}}$, which reflects the double
weight of the point measures at each vertex in this limit. For angles
$\varphi$ with $\tan(\varphi)$ irrational, one has extinctions
precisely for all wave vectors $k=\bigl(\frac{m^{}_{1}}{a^{}_{\varphi}},
\frac{m^{}_{2}}{a^{}_{\varphi}}\bigr)$ with $m^{}_{1} m^{}_{2}=0$ and
$m^{}_{1}+m^{}_{2}\in 2\mathbb{Z}+1$.  
When $\tan(\varphi)$ is rational, there are further extinctions,
which can be calculated from the explicit formula for
the diffraction measure $\widehat{\gamma^{}_{\varphi}}$.

The diffraction patterns for the two examples ($\varphi=\pi/8$ and
$\varphi=\pi/12$) from Figure~\ref{fig:sigma} are illustrated in
Figure~\ref{fig:sigmadiff}. A Bragg peak is represented by a dot that
is centred at the peak position and that has an area proportional to
the intensity. This choice resembles the experimental situation in a
reasonable way. Both patterns are non-periodic, due to the
incommensurate positions of the points in the fundamental cell. While
all Bragg peaks are located at positions of the corresponding dual
lattices, there is an apparent approximate $8$- or $12$-fold symmetry
in the patterns (sometimes called pseudo-symmetry), which is why we
chose these examples.

\section{Diffraction of mathematical quasicrystals}
\label{sec:qc}

We now leave the realm of lattice periodic systems to discuss
aperiodically ordered structures, in particular quasicrystals.  Before
we move on to structures with non-crystallographic symmetries, let us
briefly consider the inclusion of incommensurability in a lattice
periodic system, which can be seen as a first step towards the
structure of mathematical quasicrystals.

\subsection{Incommensurate phases}

The systematic investigation of incommensurate systems was pioneered
by de Wolff\cite{dWol74} and by Janner and Janssen.\cite{JJ77} We
refer to a recent monograph by van Smaalen\cite{SmaBook} and
references contained therein for details and background, and
concentrate on a couple of elementary examples here.

The simplest incommensurate structure arises from combining two periodic
Dirac combs with incommensurate periods, such as
\[
    \omega^{}_{\alpha} \, := \,
    \delta^{}_{\mathbb{Z}} + \delta^{}_{\alpha \mathbb{Z}} 
\]
with $\alpha>0$ irrational. While this is unphysical in the sense that
positions of scatterers become arbitrarily close, it is instructive to
look at the diffraction for this toy model. Observe the Eberlein
convolutions $\delta^{}_{\mathbb{Z}} \circledast
\delta^{}_{\alpha\mathbb{Z}} = \frac{1}{\alpha} \lambda$, which is a
consequence of $\alpha$ being irrational, and $\delta^{}_{\alpha
  \mathbb{Z}} \circledast \delta^{}_{\alpha\mathbb{Z}} =
\frac{1}{\alpha} \delta^{}_{\alpha\mathbb{Z}}$, which follows from a
simple density calculation. Then, the autocorrelation turns out to be
\[
   \gamma^{}_{\alpha} \, = \,
   \delta^{}_{\mathbb{Z}} + \frac{1}{\alpha}\, \delta^{}_{\alpha \mathbb{Z}}
   + \frac{2}{\alpha}\, \lambda \, ,
\]
which leads to the diffraction measure
\[
   \widehat{\gamma^{}_{\alpha}} \, = \,
   \delta^{}_{\mathbb{Z}} + \frac{1}{\alpha^{2}} \,
   \delta^{}_{\mathbb{Z}/\alpha} + \frac{2}{\alpha} \,
   \delta^{}_{0} 
\]
by an application of the PSF together with $\widehat{\lambda} =
\delta^{}_{0}$.  This pure point diffraction measure reflects the two
periodic constituents. There are Bragg peaks on the integer lattice
(with intensity $1$) and on the reciprocal lattice $\mathbb{Z}/\alpha$
of the lattice $\alpha \mathbb{Z}$, with intensity $\alpha^{-2}$. Note
that the intensity of the central peak is $1 + \alpha^{-2} +
2\alpha^{-1} = (1+\alpha^{-1})^2$, in line with the density of the
underlying point set. One might expect that the relative position of
the two constituent lattices does not matter, which indeed is the
case. Introducing a relative shift $u$ between the two periodic combs
does not affect the result, in the sense that the diffraction of the
Dirac comb $\omega^{}_{\alpha,u} = \delta^{}_{\mathbb{Z}} +
\delta^{}_{u+\alpha\mathbb{Z}}$ is still given by
$\widehat{\gamma^{}_{\alpha}}$, independently of the value of $u$.

\begin{figure}[t]
\centering
\includegraphics[width=0.8\columnwidth]{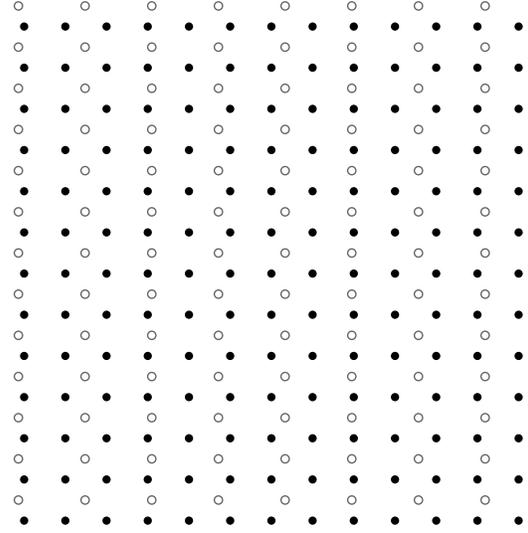}
\caption{~Composite structure comprising atoms on the square lattice
  (black dots) and on the shifted lattice $u+\varGamma$ (circles),
  with shift $u=(\frac{1}{3},\frac{1}{2})$ and lattice $\varGamma =
  \alpha\mathbb{Z} \times \mathbb{Z}$ for $\alpha=
  \tau=\frac{1}{2}(1+\sqrt{5})$.}
  \label{fig:composite}
\end{figure}

While this system is of limited practical relevance in one dimension,
one can build higher-dimensional systems using the same idea. This
results in incommensurate systems which are called \emph{composite}
structures.  Let us discuss a simple example. Fix some
$\alpha > 0$ and consider the Dirac comb
\[
   \omega \, = \, \delta^{}_{\mathbb{Z}^{2}} + \delta^{}_{u + \varGamma}
   \, = \, \delta^{}_{\mathbb{Z}^{2}} + 
   \delta^{}_{u} \ast \delta^{}_{\varGamma} \, ,
\] 
where $\varGamma = \alpha \mathbb{Z} \times
\mathbb{Z} \subset \mathbb{R}^{2}$ is a planar lattice, and $u \in
\mathbb{R}^{2}$ an arbitrary shift. For $\alpha \in \mathbb{Q}$, the
underlying point set is crystallographic, with $\mathbb{Z}^{2}
\cap \varGamma$ as its lattice of periods.  Here, we are interested
in the non-periodic case, so let us assume that $\alpha$ is irrational.
An example is displayed in Figure~\ref{fig:composite}.

\begin{figure}[t]
\centering
\includegraphics[width=0.8\columnwidth]{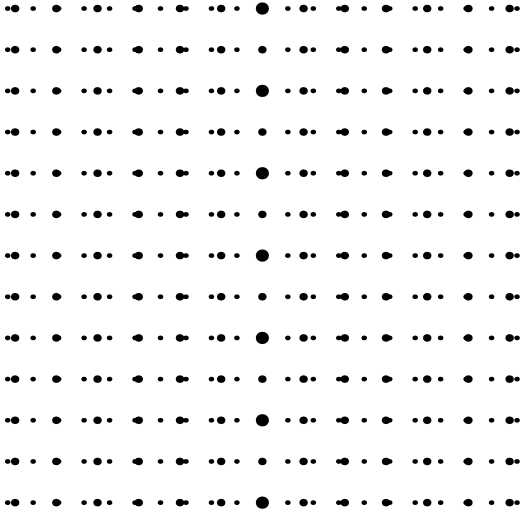}
  \caption{~Diffraction pattern of the composite structure of
    Figure~\ref{fig:composite}. Each Bragg peak is again represented by a
    dot which is centred at the position of the peak and whose area is
    proportional to the intensity. One can clearly recognise the peaks
    on the two lattices $\mathbb{Z}$ and $\varGamma^{*}$, and the
    alternating intensity of the peaks along the vertical axis, which
    are due to the choice $u^{}_{2}=\frac{1}{2}$.}
  \label{fig:compdiff}
\end{figure}

The autocorrelation for the Dirac comb $\omega$ evaluates as
\[
   \gamma \, = \, \delta^{}_{\mathbb{Z}^{2}} + \frac{1}{\alpha} \,
      \delta^{}_{\varGamma} + \frac{1}{\alpha} \,
      (\delta^{}_{u} + \delta^{}_{-u}) \ast 
      (\lambda \otimes \delta^{}_{\mathbb{Z}}) \, ,
\]
where $\mu \otimes \nu$ stands for the (tensor) product of two
measures.  The Fourier transform of $\gamma$ can be obtained by
applying the Poisson summation formula and the convolution theorem. It
has the form
\[
   \widehat{\gamma} \, = \, \delta^{}_{\mathbb{Z}^{2}}
    + \frac{1}{\alpha^{2}} \, \delta^{}_{\varGamma^{*}} + 
     \frac{2}{\alpha} \, \cos (2 \pi k^{}_{2} u^{}_{2} )
     \, (\delta^{}_{0} \otimes \delta^{}_{\mathbb{Z}})
\]
with the dual (reciprocal) lattice
$\varGamma^{*}=(\frac{1}{\alpha}\mathbb{Z})\times\mathbb{Z}$.  Note
that the final term only involves the second components of $k$ and
$u$, due to the presence of the term $\delta^{}_{0}$ in the measure
(so only $k^{}_{1}=0$ contributes). In the diffraction measure, the
composite structure is visible via additional intensities
of the peaks along the vertical axis. The total intensity of a
Bragg peak at position $(0,n)$ with $n\in\mathbb{Z}$ is
\[
   \widehat{\gamma} \bigl( \{ (0,n)\}\bigr) \, = \,
   1 + \frac{1}{\alpha^{2}} + \frac{2}{\alpha} \,
   \cos (2 \pi n u^{}_{2}) \, \ge \,
   \bigl( 1 - \frac{1}{\alpha} \bigr)^{2} \, \ge \, 0\, .
\]
The corresponding diffraction pattern for the example of
Figure~\ref{fig:composite} is shown in Figure~\ref{fig:compdiff}. 

Of course, this is merely a sketch of any real system. For a more
realistic system, one should take into account the modulation in the
positions induced by the different local 
neighbourhoods.\cite{SmaBook,Wel04,SW,WS}

Here, we consider a simpler case, based on the modulation of a
periodic structure.  A \emph{modulated structure} arises by locally
displacing positions of a crystalline point set, ensuring a minimal
distance between the new positions.  For example, start with the
integer lattice $\mathbb{Z}$ and deform it by moving the points
according to a real-valued displacement function $h$. The deformed
point set is then given by
\begin{equation}\label{eq:def-incomm}
     \varLambda^{}_{h} \, = \, \{ n + h(n) \mid n\in\mathbb{Z} \}\, ,
\end{equation}and $\delta^{}_{\varLambda_{h}}$ denotes the corresponding Dirac
comb.  To be concrete, consider the displacement function $h(n) =
\varepsilon \{ \alpha n\}$, where $\alpha$ and $\varepsilon$ are real
numbers and where $\{ x \} = x - [x]$ denotes the fractional part of
$x$. Since $|h(n)|\le \varepsilon$, the deformed point set respects a
minimum distance between points, as long as $\varepsilon$ is
sufficiently small. Clearly, if $\alpha$ is a rational number, the
resulting point set is once again periodic, while it is non-periodic for
irrational values of $\alpha$, which is the case we are interested in
here.

\begin{figure}[t]
\centering
\includegraphics[width=0.8\columnwidth]{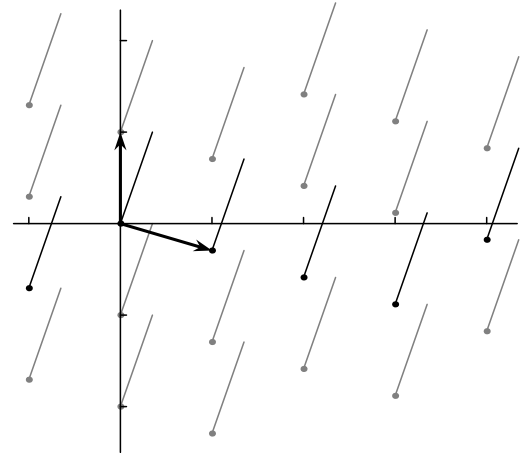}
\caption{~Superspace approach for the modulated point set
  $\varLambda^{}_{h}$ of Eq.~\eqref{eq:def-incomm}, for
  $\varepsilon=0.35$ and $\alpha\approx 0.2941$.  The lines (or
  `targets') intersecting the horizontal axis are shown in black.}
  \label{fig:modulated}
\end{figure}

To understand the corresponding set $\varLambda^{}_{h}$, it is
advantageous to use an embedding in the plane, known as the
`superspace approach' in crystallography.\cite{SmaBook} Define a
planar lattice as the integer span of two basis vectors
\[
    \varGamma = \left\langle  \begin{pmatrix} 1 \\
    -\alpha \end{pmatrix} ,  \begin{pmatrix} 0 \\ 1 \end{pmatrix} 
    \right\rangle_{\mathbb{Z}}\, ,
\]
where we use the notation $\langle u,v \rangle^{}_{\mathbb{Z}} = \{ m
u + n v \mid m,n \in \mathbb{Z} \}$. Consider now the line pattern
obtained as the $\varGamma$-orbit of the line from the origin to the
point $(\varepsilon, 1)$ (with the end point not included). Then,
$\varLambda^{}_{h}$ is the set of intersections of the horizontal axis
with these line segments; see Figure~\ref{fig:modulated} for an
illustration.

Using the fact that, for irrational $\alpha$, the sequence of numbers
$(\{\alpha n \})^{}_{n\in\mathbb{Z}}$ is uniformly distributed in the
unit interval,\cite{KNBook} one can calculate the autocorrelation
$\gamma^{}_{h}$ of the Dirac comb on $\varLambda^{}_{h}$
explicitly. The result is
\[
    \gamma^{}_{h} \, = 
    \sum_{m\in\mathbb{Z}} \bigl( (1 - \{ \alpha m\} ) \,
    \delta^{}_{m + \varepsilon \{ \alpha m \}} + \{ \alpha m \}\,
    \delta^{}_{m - \varepsilon ( 1 - \{ \alpha m \})} \bigr) .
\]
The corresponding diffraction measure $\widehat{\gamma^{}_{h}}$ 
reads
\begin{equation}\label{eq:moddiff}
    \widehat{\gamma^{}_{h}} \; = \, \sum_{k \in \mathbb{Z}[\alpha]}
    \lvert A (k) \rvert^{2} \, \delta^{}_{k}  ,
\end{equation}
with (complex) amplitudes 
\begin{equation}\label{eq:modamp}
    A (k)\, = \, e^{- \pi i k^{\star}} \,
   \mathrm{sinc} (\pi k^{\star}) \, ,
\end{equation}
where $\mathrm{sinc} (x) = \sin (x) / x$.  The map $k\mapsto
k^{\star}$ acts on elements of $\mathbb{Z}[\alpha]=\{r+s\alpha\mid
r,s\in\mathbb{Z}\}$ as $(r + s \alpha) \mapsto \bigl(r \varepsilon + s
(1 + \varepsilon \alpha)\bigr)$ for any $r,s \in \mathbb{Z}$. In this
example, $\widehat{ \gamma^{}_{h}}$ is a pure point measure which is
supported on a dense set. Despite the denseness of the Bragg peaks,
the total intensity scattered into any compact subset of $\mathbb{R}$
is finite, because the intensities are locally summable. The proof for
the diffraction formula is non-trivial. However, this can be
interpreted as a special case of the diffraction of model sets (cut
and project sets), because the modulated structure
\eqref{eq:def-incomm} is in fact a model set. We now turn our
attention to this general notion, and discuss a number of relevant
examples and their diffraction.

\subsection{Model sets}

There are a number of ways to construct aperiodically ordered
systems.\cite{GS02} From the viewpoint of diffraction, the best
understood is a natural generalisation of lattice-periodic structures
obtained by a projection from a higher-dimensional lattice. Such
systems are called \emph{cut and project sets} or \emph{model
  sets},\cite{M} and can be produced in a number of essentially
equivalent ways,\cite{GR86} including de Bruijn's grid
method\cite{deBr81} and Kramer's `Klotz construction',\cite{KS89} as
well as a number of other approaches.\cite{MP96,HKPM97}

The model set approach can be viewed as a generalisation of the notion
of a quasiperiodic function.\cite{Bohr} In the simplest setting, the
idea is much like what we saw for the modulated phase in
Figure~\ref{fig:modulated} above: The aperiodic structure emerges by
taking a cut across a higher-dimensional periodic structure, using a
direction that is incommensurate with the lattice. The general setting
for the case of Euclidean model sets is encoded in the \emph{cut and
  project scheme} (CPS)
\begin{equation}\label{eq:cps}
\renewcommand{\arraystretch}{1.2}\begin{array}{r@{}ccccc@{}l}
   & \mathbb{R}^{d} & \xleftarrow{\,\;\;\pi\;\;\,} 
         & \mathbb{R}^{d} \times \, \mathbb{R}^{m}\!  & 
        \xrightarrow{\;\pi^{}_{\mathrm{int}\;}} & \mathbb{R}^{m} & \\
   & \cup & & \cup & & \cup & \hspace*{-2ex} 
   \raisebox{1pt}{\text{\footnotesize dense}} \\
   & \pi(\mathcal{L}) & \xleftarrow{\; 1-1 \;} & \mathcal{L} & 
   \xrightarrow{\; \hphantom{1-1} \;} & 
       \pi^{}_{\mathrm{int}}(\mathcal{L}) & \\
   & \| & & & & \| & \\
   &  L & \multicolumn{3}{c}{\xrightarrow{\qquad\quad\quad \;\;
       \;\star\; \;\; \quad\quad\qquad}} 
       &   {L_{}}^{\star} & \\
\end{array}\renewcommand{\arraystretch}{1}
\end{equation}
where $\mathbb{R}^{d}$ is the physical (sometimes also called direct or
parallel) space, and $\mathbb{R}^{m}$ is referred to as the internal
(or perpendicular) space. Here, $\mathcal{L}\subset \mathbb{R}^{d+m}$
is a lattice in $d+m$ dimensions, and $\pi$ and
$\pi^{}_{\mathrm{int}}$ denote the natural projections onto the
physical and internal spaces.  It is assumed that
$L=\pi(\mathcal{L})\subset\mathbb{R}^{d}$ is a bijective image of
$\mathcal{L}$ in direct space, and that the set
$L^{\star}=\pi^{}_{\mathrm{int}}(\mathcal{L})\subset\mathbb{R}^{m}$ is
dense in internal space. As a consequence, the $\star\,$-map\cite{M}
$x\mapsto x^{\star}$ is well-defined on $L$.

A \emph{model set} for a given CPS is then a set of the form
\begin{equation}\label{eq:ms}
    \varLambda \, = \,
    \bigl\{  x\in L \mid  x^{\star} \in W \bigr\} ,
\end{equation}
where $W\subset\mathbb{R}^{m}$ (called the \emph{window} or
\emph{acceptance domain}) is a relatively compact subset of
$\mathbb{R}^{m}$ with non-empty interior. More generally, also
translates of such sets are called model sets. The elements of the
model set $\varLambda$ lie in the projected lattice $L$ in direct
space, and the window in internal space determines which elements of
$L$ are selected. The conditions on the window ensure that the model
set $\varLambda$ is a Delone set.  In fact, a model set $\varLambda$
is always a Meyer set,\cite{Meyer,M} which means that
$\varLambda-\varLambda:=\{x-y\mid x,y\in\varLambda\}$ is uniformly
discrete, while $\varLambda$ is relatively dense. Note that uniform
discreteness of $\varLambda - \varLambda$ implies that of $\varLambda$,
and is actually a \emph{much} stronger condition.\cite{Meyer,M,Lag,Lag99}

Clearly, the projection approach produces point sets in space rather
than the tilings that are conventionally used to model atomic
structures of quasicrystals. However, as long as there exists a
\emph{local} rule to switch from the point set to the tiling picture
\emph{and} vice versa, we can consider both structures as equivalent
(as any atomic structure will be a local decoration of either), or
shortly as MLD (which stands for mutual local derivability).\cite{B} For
instance, in one dimension, a tiling of $\mathbb{R}$ by two intervals
of different lengths is clearly MLD with the set of left endpoints of
all intervals.

In what follows, we only consider \emph{regular model sets}, so we
require that the boundary $\partial W$ of the window $W$ has zero
Lebesgue measure. The Euclidean setting \eqref{eq:cps} generalises to
the case where the internal space is a locally compact Abelian
group.\cite{Meyer,M,Martin} We shall meet an example later, where the
internal space is based on $2$-adic integers.

Regular model sets are pure point diffractive,\cite{Hof,Martin,Crelle}
and in this sense are natural generalisations of lattices. This is a
central result of the theory of model sets which has been proved by
methods of dynamical systems theory,\cite{Hof,Martin,LS} in terms of
almost periodic measures\cite{Crelle,MS04,S05} and, following a
suggestion by Lagarias, by using the Poisson summation formula for the
embedding lattice and Weyl's lemma on uniform distribution.\cite{TAO}
The diffraction measure $\widehat{\gamma}$ of the Dirac comb
$\delta^{}_{\varLambda}$ is explicitly given by
\begin{equation}\label{eq:modeldiff}
    \widehat{\gamma}\,  = \sum_{k\in L{}_{}^{\circledast}}
         \lvert A(k) \rvert^{2}\, \delta_{k}\, .
\end{equation}
Here, $L^{\circledast} = \pi (\mathcal{L}^{*})$ is the corresponding Fourier
module, which is the projection of the higher-dimensional dual lattice.
The amplitudes are given be 
\begin{equation}\label{eq:modelamp}
   A(k) \, = \, 
  \frac{\mathrm{dens} (\varLambda)}{\mathrm{vol} (W)}
  \, \widehat{1^{}_{\! W}} (-k^{\star})\, ,
\end{equation}
where $1^{}_{W}$ is the characteristic function of the window
$W$. Various generalisations, in particular to certain weighted Dirac
combs, have been discussed in the
literature.\cite{Martin,Crelle,TAO,Richard} An alternative (and
somewhat complementary) approach based on an average periodic
structure can be employed to unravel various modulation features in
the diffraction patterns of quasicrystals. This is systematically
explained in a recent review\cite{Wolny} by Wolny and coworkers; see
references cited there for further details.

\subsection{One-dimensional examples}

\begin{figure}[t]
\centering
\includegraphics[width=0.8\columnwidth]{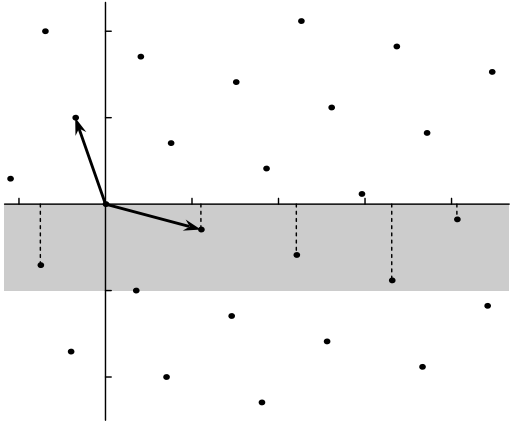}
  \caption{~Model set description of the 
  modulated point set $\varLambda_{h}$ of
  Figure~\ref{fig:modulated}.}
  \label{fig:modcps}
\end{figure}

We start by re-expressing the modulated point set $\varLambda_{h}$ of
Eq.~\eqref{eq:ms} as a cut and project set. To this end, we need to
write $\varLambda_{h}$ via an orthogonal projection, rather than via the
(implicit) skew projection of Figure~\ref{fig:modulated}. This can be
done by introducing the matrix $A = \left(
  \begin{smallmatrix} 1 & -\varepsilon \\ 0 & 1 \end{smallmatrix}
\right)$ and considering the lattice $\mathcal{L} = A \varGamma$. This
lattice and its dual lattice are given in terms of a
$\mathbb{Z}$-basis by
\[
   \mathcal{L}  =\,
   \bigg\langle\! \binom{1 + \varepsilon \alpha}{-\alpha} , 
   \binom{-\varepsilon}{1}\!\bigg\rangle_{\!\!\mathbb{Z}},\quad
    \mathcal{L}^{*}  =  \,\bigg\langle\!
    \binom{1}{\varepsilon}, \binom{\alpha}{1+\varepsilon \alpha}
    \!\bigg\rangle_{\!\!\mathbb{Z}} .
\]
The two generating vectors and the lattice points of $\mathcal{L}$ are
shown in Figure~\ref{fig:modcps}.

The set $\varLambda_{h}$ is now a model set for the CPS with lattice
$\mathcal{L}\subset\mathbb{R}^{2}=\mathbb{R}\times\mathbb{R}$ (so
$d=m=1$ and both direct and internal space are $\mathbb{R}$). The
window is the interval $W=[0,-1)$, and the condition $x^{\star}\in W$
selects all lattice points that are located within the shaded strip of
Figure~\ref{fig:modcps} (which is the reason why this approach is
sometimes also referred to as the strip projection method). For
$\varepsilon = 0$, we get a (non-minimal) embedding of $\mathbb{Z}$ in
$\mathbb{R}^{2}$, and for rational $\alpha = \frac{p}{q}$ with coprime
integers $p$ and $q$ we obtain a periodic point set with lattice of
periods $q \mathbb{Z}$.

The formulas \eqref{eq:moddiff} for the diffraction and
\eqref{eq:modamp} for the amplitudes now follow from the general
result of Eqs.~\eqref{eq:modeldiff} and \eqref{eq:modelamp}. The
Fourier module is $L^{\circledast} = \pi (\mathcal{L}^{*}) =
\mathbb{Z}[\alpha]$, and the action of the $\star$-map can be read off
from the explicit bases of $\mathcal{L}$ and $\mathcal{L}^{*}$ given
above.\medskip

The most frequently invoked example of a one-dimensional
(mathematical) quasicrystal is the \emph{Fibonacci chain}.  Its
geometric version is built from two intervals (prototiles) $L$ and $S$
(for long and short) of lengths $\tau=(1+\sqrt{5})/2$ and $1$.  It can
be generated by iterating the square of the inflation rule $L\mapsto
LS$, $S\mapsto L$, starting from a legal seed (such as $L|L$, where
the vertical line indicates the reference point). This leads to the
two-sided interval sequence
\[
   .\!.\!.\,  LSLLSLSLLSLLSLSLLSLSL | 
           LSLLSLSLLSLLSLSLLSLSL\,
   .\!.\!.
\]
The bi-infinite sequence is aperiodic, with relative frequencies
$\tau^{-1}$ and $\tau^{-2}$ for the two prototiles. 

\begin{figure}[t]
\centering
\includegraphics[width=\columnwidth]{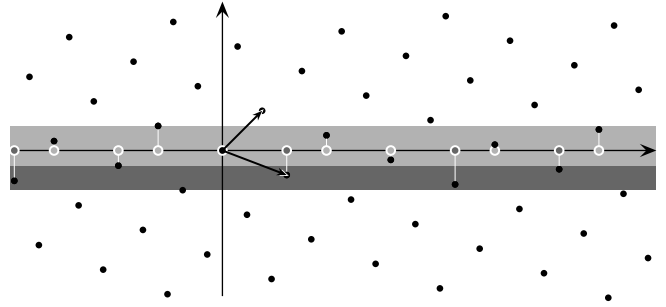}
  \caption{~Model set description of the Fibonacci chain.}
  \label{fig:fibocps}
\end{figure}

Define two point sets $\varLambda_{L}$ and $\varLambda_{S}$ as the
left endpoints of the corresponding intervals in the chain, taking the
reference point as $0$.  They are model sets for the CPS
\eqref{eq:cps} with $d=m=1$ and
\[
   L \, = \, \mathbb{Z}[\tau] \, = \,
   \{m+n\tau\mid m,n\in\mathbb{Z}\}\, .
\]
The corresponding planar lattice is
\[
   \mathcal{L} \, = \, \bigg\langle\!
   \binom{1}{1}, \binom{\tau}{1-\tau} 
   \!\bigg\rangle_{\!\!\mathbb{Z}}\, , 
\]
which has density $1/\sqrt{5}$ and the dual lattice
\[
   \mathcal{L}^{*} \, = \, \frac{2\tau-1}{5}\bigg\langle\!
   \binom{\tau-1}{\tau}, \binom{1}{-1} 
   \!\bigg\rangle_{\!\!\mathbb{Z}}\, .
\]
One has $\varLambda_{L,S}=\{x\in L\mid x^{\star}\in W_{L,S}\}$
with the windows
\[
    W_{L} \, = \, (-1,\tau-2]\quad\text{and}\quad
    W_{S} \, = \, (\tau-2,\tau-1]
\]
and the $\star$-map defined by $\sqrt{5}\mapsto -\sqrt{5}$, so that
$(m+n\tau)^{\star}=m+n-n\tau$. The construction is illustrated in
Figure~\ref{fig:fibocps}. The Fibonacci model set is $\varLambda =
\varLambda_{L} \cup\varLambda_{S}$, with window 
\[
   W \, = \, W_{L}\cup W_{S} \, =\,  (-1,\tau-1]\, .
\]
This way, $\varLambda$ is a point set of density $\tau/\sqrt{5} =
(\tau + 2)/5$.  Note that it is possible to modify the embedding
lattice $\mathcal{L}$ by scaling the internal space relative to
physical space. In particular, multiplying the scale of internal space
by $\tau$, the embedding lattice is a rotated copy of
$\sqrt{\tau+2}\;\mathbb{Z}^{2}$.

\begin{figure}[t]
\centering
\includegraphics[width=0.8\columnwidth]{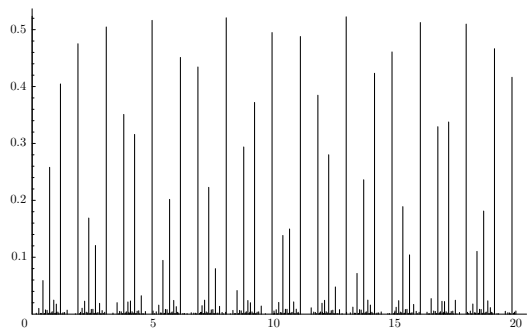}
\caption{~Diffraction pattern for the Fibonacci chain $\varLambda$.
  The Bragg peak at $0$ has height $(\mathrm{dens} (\varLambda))^{2} =
  (\tau+1)/5\approx 0.5206$, and the entire pattern is reflection
  symmetric.}
  \label{fig:fibodiff}
\end{figure}

The Dirac comb $\omega=\delta_{\varLambda}$ is pure point diffractive,
by an application of the model set diffraction
theorem\cite{Hof,Martin,Crelle} mentioned before. The diffraction
measure $\widehat{\gamma}$ is explicitly given by
Eq.~\eqref{eq:modeldiff} with the amplitudes
\begin{equation}
   A(k) \, = \, e^{\pi i k^{\star} (\tau-2)}\,\frac{\tau + 2}{5}\, 
   \mathrm{sinc}(\pi \tau k^{\star})  
\end{equation}
via Eq.~\eqref{eq:modelamp}, where $\mathrm{sinc}(x)=\sin(x)/x$. The
phase factor reflects the position of the window, which is centred at
$(\tau-2)/2$. The sum in Eq.~\eqref{eq:modeldiff} runs over the 
Fourier module
\[
     L^{\circledast} \, =\, \pi (\mathcal{L}^{*}) \, = \,
     \frac{1}{\sqrt{5}}\, \mathbb{Z}[\tau] \, .
\]
A sketch of the diffraction pattern is shown in
Figure~\ref{fig:fibodiff}. Note that the intensity function
$I(k)=\lvert A(k)\rvert^{2}$ vanishes if and only if $\tau
k^{\star}\in \mathbb{Z}\setminus\{0\}$. This means $k=\ell\tau$ with
$0\ne \ell\in\mathbb{Z}$. Since all such points lie in the Fourier
module $L^{\circledast}$, we have identified all extinctions. These
are a fingerprint of the intrinsic inflation symmetry.\medskip

As an example of a limit-periodic structure, consider the \emph{period
doubling sequence}. Written as an element $w\in\{0,1\}^{\mathbb{Z}}$,
it is given by $w(2n)=1$, $w(4n+1)=0$ and $w(4n+3)=w(n)$ for
$n\in\mathbb{Z}$. This rule specifies every position except $n=-1$,
where we can choose either possibility.  Both possibilities can also
be obtained as a fixed point sequence of the square of the
substitution $1\mapsto 10$, $0\mapsto 11$. The two sequences have cores
\[
  \ldots 101110101011101 \hspace{0.5pt}
  \raisebox{-3pt}{$\stackrel{\scriptstyle 1}{\scriptstyle 0}$}\,  
  |1011101010111011 \ldots 
\]
and are
locally indistinguishable. They thus define the same system.  The
underlying Toeplitz structure of a hierarchy of scaled and shifted
copies of $\mathbb{Z}$ is apparent from the
formula\cite{BMS,Crelle,TAO}
\[
     \varLambda \, = \, \{n\in\mathbb{Z}\mid w(n)=1\} \, = \,
      \bigcup_{\ell\ge 0} \bigl(( 2\cdot 
      4^{\ell}\mathbb{Z} + (4^{\ell}-1)\bigr) 
\]
for $w(-1)=0$ (with $-1$ added to $\varLambda$ in the other case). 
This set can be described as a model set, but with the internal space
being the $2$-adic integers. Consequently, the diffraction measure
of the Dirac comb $\delta_{\varLambda}$ is again pure point.

\begin{figure}[t]
\centering
\includegraphics[width=0.8\columnwidth]{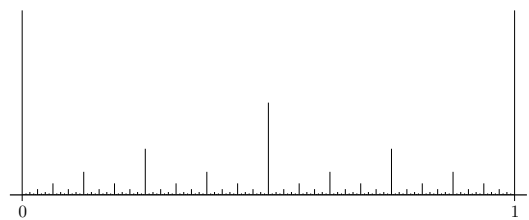}
  \caption{~Absolute values of the diffraction amplitudes 
   for the period doubling chain. The diffraction pattern is
   $1$-periodic.}
  \label{fig:pddiff}
\end{figure}

The corresponding diffraction formula can be given explicitly as
follows.\cite{BG11,TAO} The Fourier module is 
\[
    L^{\circledast}\, = \, \mathbb{Z}[\tfrac{1}{2}] \, = \,
    \bigl\{ \tfrac{m}{2^{r}}\mid (r=0,m\in\mathbb{Z})
    \text{ or } (r\ge 1, m\text{ odd})\bigr\}\, ,
\]
so that we can again use Eq.~\eqref{eq:modeldiff}. Here, the amplitudes
are
\[
    A(k) \, = \, \frac{2}{3}\, \frac{(-1)^{r}}{2^{r}}\, 
    e^{2\pi ik}\, ,
\]
with $k=\frac{m}{2^r}\in L^{\circledast}$. This parametrisation
specifies $k$ uniquely. Figure~\ref{fig:pddiff} shows the absolute
values $\lvert A(k)\rvert$ for $k\in L^{\circledast}\cap [0,1]$.  This
pattern repeats $\mathbb{Z}$-periodically.

Further one-dimensional examples will be discussed in
Section~\ref{sec:cont} in the context of continuous diffraction
measures. Let us now turn our attention to higher-dimensional model
sets.

\subsection{Cyclotomic model sets}

For the description of two-dimensional tilings, it is advantageous to
work with complex numbers $x+iy$ in $\mathbb{C}$ rather than with
points $(x,y)$ in $\mathbb{R}^{2}$. In $\mathbb{C}$, a rotation by an
angle $\varphi$ just corresponds to multiplication with the complex
number $e^{i\varphi}$. This point of view is a natural generalisation
of de Bruijn's method\cite{deBr81} and the Fourier space
approach.\cite{MRW87} A natural way to implement an $n$-fold
rotational symmetry is to choose a primitive $n$th root of unity
$\xi^{}_{n}\in\mathbb{C}$ (so $\xi_{n}^{n}=1$ and $\xi_{n}^{m}\ne 1$
for $1\le m<n$), and to consider the $\mathbb{Z}$-module
$\mathbb{Z}[\xi^{}_{n}]$ of \emph{cyclotomic integers}, comprising all
integer linear combinations of powers of $\xi^{}_{n}$ (the solutions of
the equation $x^{n}=1$). One can think of cyclotomic integers as the
set of all points in the plane that can be reached by taking steps of
unit length along the directions of a regular $n$-star.  Clearly, the
resulting point set is symmetric under rotations of multiples of
$2\pi/n$; in fact, under rotations by multiples of $\pi/n$ if $n$ is
odd. Therefore, one usually restricts to integers $n\not\equiv 2\bmod
4$ to avoid duplications.

The cases $n\in\{1,2\}$ are trivial in the sense that the resulting
point sets lie on the real axis. The choices $n\in\{3,4\}$ lead to
crystallographic point sets, the triangular lattice with sixfold
symmetry and the square lattice with fourfold symmetry. Any other
choice $n \ge 5$, $n\not\equiv 2\bmod 4$, produces a dense point set
in the plane, with $n$-fold symmetry for even $n$, and $2n$-fold
symmetry for odd $n$.

The dense point set $\mathbb{Z}[\xi^{}_{n}]$ can be embedded into a
lattice by lifting it to a suitable higher-dimensional space,
essentially by making all directions in the $n$-star that are linearly
independent over the integers (there are $\phi(n)$ such directions,
where $\phi$ is Euler's totient function) also linearly independent
over the real numbers. A natural way to do this is the Minkowski (or
Galois) embedding
\begin{equation}\label{eq:cycloembed}
    \mathcal{L}_{n}\, =\, \bigl\{ (x,\sigma^{}_{2}(x),\ldots,
                       \sigma^{}_{\frac{1}{2}\phi(n)}(x))
              \;\big|\; x\in\mathbb{Z}[\xi^{}_{n}]\bigr\}
\end{equation}
which defines a lattice $ \mathcal{L}_{n}\subset
\mathbb{C}^{\frac{1}{2}\phi(n)}_{} \simeq
\mathbb{R}^{\phi(n)}_{}$. Here, $\sigma^{}_{\ell}$, with
$1\le\ell\le\phi(n)$, are the Galois automorphisms of the
corresponding cyclotomic number field, mapping a primitive root
$\xi^{}_{n}\mapsto\xi^{m^{}_{\ell}}_{n}$ to a primitive root
$\xi^{m^{}_{\ell}}_{n}$, where $\{m^{}_{\ell}\mid
1\le\ell\le\phi(n)\}= \{1\le k\le n\mid \text{$k$ and $n$ coprime}\}$,
together with a suitable ordering.  Note that $\sigma^{}_{1}$ is the
identity map. Using the lattice $\mathcal{L}_{n}$ in a cut and project
scheme, with physical space $\mathbb{R}^{2}\simeq\mathbb{C}$ and
internal space $\mathbb{R}^{\phi(n)-2}_{}$, we produce
\emph{cyclotomic model sets}, which, for suitably chosen windows, have
$n$-fold ($2n$-fold) rotational symmetry.

As an explicit example, we consider the classic Ammann-Beenker (or
octagonal) tiling\cite{AGS,Beenker} as a cyclotomic model set with
$n=8$. Other standard examples of this type include the Penrose
tiling\cite{Pen74} of Figure~\ref{fig:pen} and the T\"{u}bingen
triangle tiling\cite{BKSZ90} (both with tenfold symmetry) and
G\"{a}hler's shield tiling\cite{GaehlerDiss,Gaeh93} (with twelvefold
symmetry). The latter is locally equivalent (MLD) with a tiling
introduced by Socolar.\cite{Soc89} Since $\phi(5)=\phi(8)=\phi(12)=4$,
all these tilings are obtained from cut and project schemes
\eqref{eq:cps} with internal space $\mathbb{R}^{2}$.

\begin{figure}[t]
\centering
\includegraphics[width=\columnwidth]{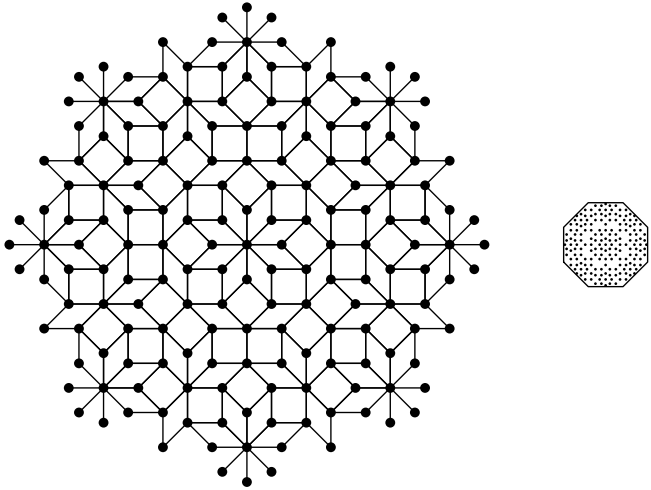}
  \caption{~Ammann-Beenker tiling as a cyclotomic model set.}
  \label{fig:abtil}
\end{figure}

Of course, the resulting tilings are only rotationally symmetric if
the window is chosen to have an appropriate rotational symmetry. To
obtain the (undecorated) Ammann-Beenker tiling, the window
$W^{}_{\mathrm{AB}}$ has to be chosen as a regular octagon, of unit
edge length. The module
\[
   L \, = \, \mathbb{Z}[\xi^{}_{8}] \, = \,
   \{n^{}_{0}+n^{}_{1}\xi_{8}^{}+n^{}_{2}\xi_{8}^{2}+
    n^{}_{3}\xi_{8}^{3}\mid
    (n^{}_{0},n^{}_{1},n^{}_{2},n^{}_{3})
    \in\mathbb{Z}^{4}\}
\]
is dense in the plane, and naturally lifts to a hypercubic lattice in
four dimensions (the corresponding Minkowski embedding $\mathcal{L}_8$
is a scaled and rotated version of $\mathbb{Z}^{4}$). The $\star$-map
can be chosen as the Galois automorphism $\xi^{}_{8}\mapsto
\xi^{3}_{8}$, and the Ammann-Beenker model set is then obtained as
\[
   \varLambda^{}_{\mathrm{AB}}\, = \, 
   \{x\in L\mid x^{\star} \in W^{}_{\mathrm{AB}} \}\, .
\]
Figure~\ref{fig:abtil} shows the picture in physical and internal
space. Selecting points $x\in L$ whose $\star$-image falls inside
the octagonal window (shown on the right of Figure~\ref{fig:abtil})
produces the point set in physical space shown on the left. Connecting
all points of unit distance (which clearly is a local rule) recovers
the Ammann-Beenker tiling, which is MLD with the cyclotomic model set.
The decorations needed for the approach via local rules\cite{AGS} add
some non-local information, and cannot be recovered from the
undecorated tiling alone.\cite{Soc89,Gaeh93}.

\begin{figure}[t]
\centering
\includegraphics[width=\columnwidth]{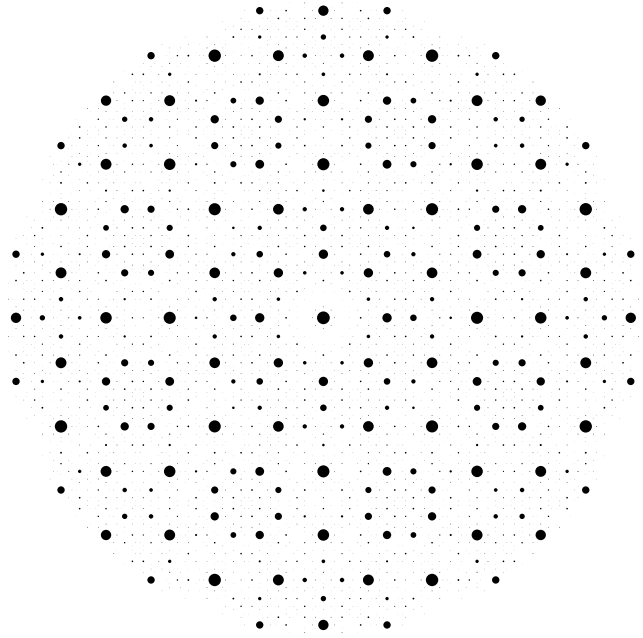}
  \caption{~Diffraction pattern of the Ammann-Beenker tiling.}
  \label{fig:abdiff}
\end{figure}

The diffraction of the Dirac comb on the Ammann-Beenker model set can
be calculated via Eqs.~\eqref{eq:modeldiff} and \eqref{eq:modelamp}.
It is a pure point measure supported on the dual module
$L^{\circledast}=\frac{1}{2} L$ (with the factor $\frac{1}{2}$ due to
the aforementioned scaling of the hypercubic lattice in the Minkowski
embedding).  The amplitudes (or Fourier-Bohr coefficients) are
\[
   A(k)\, =\, \frac{1}{4} \, 
   \widehat{1_{W^{}_{\mathrm{AB}}}}(-k^{\star})
\]
because the lattice $\mathcal{L}^{}_{8}$ has density $\frac{1}{4}$. A
central patch of the diffraction image, obtained via an exact
calculation of the Fourier transform of the octagonal window, is shown
in Figure~\ref{fig:abdiff}. In principle, the diffraction of any model
set can be calculated (at least approximately) in this way, although
it may be complicated if the window is not a simple polygon or circle,
such as for the square-triangle tilings where the windows have fractal
boundaries.\cite{BKS,HRB97}

\subsection{Icosahedral model sets}

The model set approach works in any dimension. In particular, it can
be used to construct icosahedrally symmetric tilings in
three-dimensional space, which are particularly relevant for
applications in crystallography. The minimum embedding dimension for
this purpose is six, because one needs a faithful action of the
icosahedral group and an invariant subspace of dimension $3$.  In this
setting, there exist three different classes of icosahedral model
sets, which correspond to the three different hypercubic lattices
(primitive, face-centred and body-centred) in six
dimensions.\cite{Schw,RMW87} As body-centred icosahedral structures have
not yet been identified in quasicrystals, we concentrate on the other
two classes, and discuss one example of either type.

\begin{figure}[t]
\centering
\includegraphics[width=0.6\columnwidth]{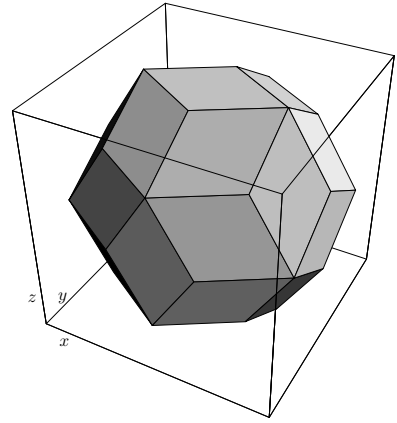}
  \caption{~Kepler's triacontahedron as window of the primitive 
  icosahedral tiling due to Kramer and Neri.\cite{KN}}
  \label{fig:knwindow}
\end{figure}

For the \emph{primitive icosahedral tiling}, we start from a lattice
$\mathcal{L}$ that is similar to the integer lattice $\mathbb{Z}^{6}$,
and use a cut and project scheme \eqref{eq:cps} where both physical
and internal space are $\mathbb{R}^{3}$. The corresponding window is
shown in Figure~\ref{fig:knwindow}; it is a semi-regular polyhedron
known as Kepler's triacontahedron. The triacontahedron has edge length
$\sqrt{2+\tau}$, volume $20\tau^{3}$ and surface area $60\tau$, where
$\tau=(1+\sqrt{5})/2$ is again the golden ratio.  This approach was
pioneered by Kramer and Neri,\cite{KN} and the tiling is also
sometimes called the Ammann-Kramer-Neri tiling (Ammann described the
tiling earlier by different means, without publishing his findings;
compare the corresponding comments in Mackay's early
paper\cite{Mac81}). Some authors also call it the three-dimensional
Penrose tiling, in analogy to the fivefold rhombus tiling in the
plane.

The primitive icosahedral tiling is built from two rhombohedral
prototiles, a thick (or prolate, called $T_{\mathrm{p}}$) and a thin
(or oblate, called $T_{\mathrm{o}}$) one. They can be defined as the
convex hulls of their vertices
\[
  \begin{split}
   T^{}_{\mathrm{p}} &=
    \mathrm{conv}\{0,v^{}_{1},v^{}_{2},v^{}_{3},v^{}_{1}+v^{}_{2},
           v^{}_{1}+v^{}_{3},v^{}_{2}+v^{}_{3},v^{}_{1}
           +v^{}_{2}+v^{}_{3}\}\, , \\
   T^{}_{\mathrm{o}} &=
    \mathrm{conv}\{0,v^{}_{1},v^{}_{2},v^{}_{5},v^{}_{1}+v^{}_{2},
           v^{}_{1}+v^{}_{5},v^{}_{2}+v^{}_{5},
           v^{}_{1}+v^{}_{2}+v^{}_{5}\} \, ,
   \end{split}
\]
where the basis vectors are\cite{RMW87}
\begin{equation}\label{eq:modbas}
 \begin{aligned}
  v^{}_{1}&=(\tau,0,1)\, ,&
  v^{}_{2}&=(\tau,0,-1)\, ,&
  v^{}_{3}&=(1,\tau,0)\, ,\\
  v^{}_{4}&=(-1,\tau, 0)\, ,&
  v^{}_{5}&=(0,1,\tau)\, ,&
  v^{}_{6}&=(0,-1,\tau)\, .
\end{aligned}
\end{equation}
These six vectors generate the primitive icosahedral module
\[
  \mathcal{M}^{}_{\textrm{P}}\, =\, \bigl\langle v^{}_{1}, v^{}_{2},
  v^{}_{3}, v^{}_{4},  v^{}_{5}, v^{}_{6}\bigr\rangle^{}_{\mathbb{Z}}\, ,
\]
which plays the role of $L=\pi(\mathcal{L})$ in the corresponding cut
and project scheme \eqref{eq:cps}.  The $\star$-map acts as
$(a,b,c)\mapsto \tau (a^{\prime},b^{\prime},c^{\prime})$ on
$\mathcal{M}^{}_{\textrm{P}}$, where ${}^{\prime}$ denotes algebraic
conjugation (which maps $\sqrt{5}\mapsto-\sqrt{5}$, hence
$\tau^{\prime}=1-\tau$). In this formulation, the embedding lattice
$\mathcal{L}=\{(x,x^{\star})\mid x\in L\}$ is similar to
$\mathbb{Z}^{6}$, and explicitly generated by the $\mathbb{Z}$-basis
$\{(v^{}_{i},v^{\star}_{i}) \mid 1\le i\le 6\}$ with the vectors from
Eq.~\eqref{eq:modbas}.  Consequently, the fundamental cell of
$\mathcal{L}$ has volume $40(4\tau+3)$, so that the density of
$\mathcal{L}$ is $(7-4\tau)/200$.

\begin{figure}[b]
\centering
\includegraphics[width=0.6\columnwidth]{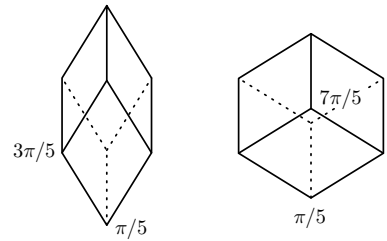}
  \caption{~Sketch of the two rhombohedral prototiles.}
  \label{fig:keprhomb}
\end{figure}

A sketch of the two prototiles is shown in Figure~\ref{fig:keprhomb}.
The rhombohedra have solid angles $\pi/5$, $3\pi/5$ and $7\pi/5$ as
indicated. The solid angles in both cases add up to $4\pi$. The
prototiles have volumes $2\tau^{2}$ (for $T_{\mathrm{p}}$) and $2\tau$
(for $T_{\mathrm{o}}$). Note that ten rhombohedra of each type can be
assembled\cite{Kow38,KN} to fill Kepler's triacontahedron of
Figure~\ref{fig:knwindow}.

\begin{figure}[t]
\centering
\includegraphics[width=0.6\columnwidth]{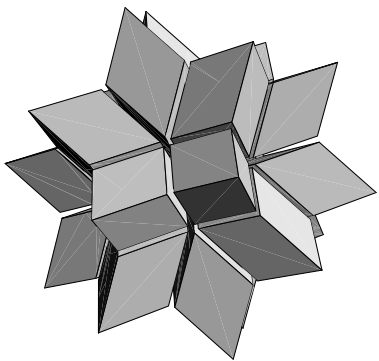}
  \caption{~Icosahedrally symmetric vertex star of the
   Kramer-Neri tiling, comprising $20$ acute rhombohedra.}
  \label{fig:knstar}
\end{figure}

Figure~\ref{fig:knstar} shows the only vertex star out of the $24$
possible vertex stars of the Kramer-Neri tiling which has
full icosahedral symmetry.  In any tiling obtained from a generic
model set, this vertex type occupies a subset that itself is a model
set with the $\tau^{-3}$-scaled triacontahedron as its window.  This
property corresponds to the invariance of the module
$\mathcal{M}_{\mathrm{p}}$ under multiplication by $\tau^{3}$ and
reflects the inflation symmetry of the primitive icosahedral tiling.
The corresponding (local) inflation rule, however, turns out to be
rather complicated and has never been presented in complete detail.

The diffraction of the Dirac comb on the primitive icosahedral
model set can be calculated again by Eqs.~\eqref{eq:modeldiff} and
\eqref{eq:modelamp}. The Fourier module in this case is
\[
    L_{}^{\circledast} \, = \, 
    \mathcal{M}^{\circledast}_{\mathrm{P}} \, = \,
   \frac{1}{2(\tau+2)} \, \mathcal{M}^{}_{\mathrm{P}}\, .
\]
The diffraction spectrum consists of a dense set of Bragg peaks
located on $L_{}^{\circledast}$, of which only a discrete subset has
intensity above any chosen (positive) threshold.  A full calculation
of the Fourier transform of the triacontahedron was given by
Elser,\cite{Els86} so the intensities can be obtained explicitly.

\begin{figure}
\centering
\includegraphics[width=0.785\columnwidth]{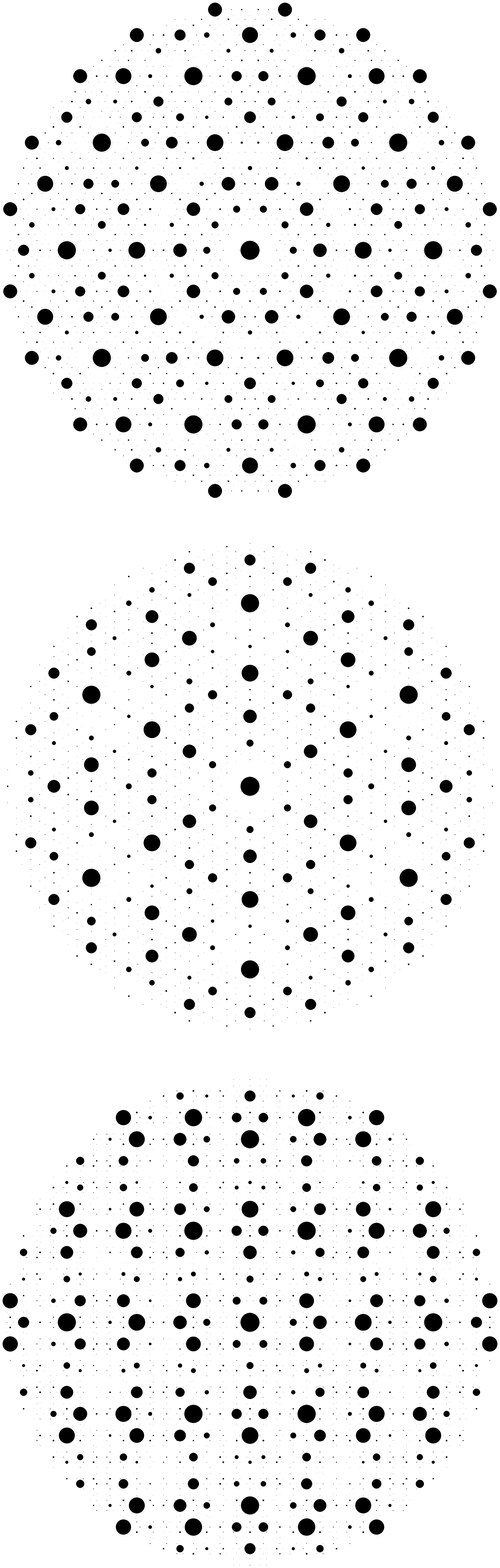}
  \caption{~Fivefold (top), threefold (middle) and twofold (bottom) sections
   of the diffraction pattern of the primitive icosahedral tiling.}
  \label{fig:primicodiff}
\end{figure}

For simplicity, however, we employ a spherical approximation to the
amplitudes, by replacing the triacontahedron by a sphere of equal
volume $20\tau^{3}$.  The radius of the sphere turns out to be
\[
   R \, = \, \Bigl(\frac{15}{\pi}\Bigr)^{1/3} \tau
     \, \approx\, 2.7246\, .
\]
Because the triacontahedral window is well approximated by this
sphere, the difference between the approximate and the exact
diffraction intensities is tiny, and irrelevant for our purpose. Note
that the approximation only affects the values of the amplitudes, not
the location of the peaks (except for extinctions, which might show up
in the approximation as tiny intensities).  The Fourier transform of
the spherical window evaluates as
\[
    \frac{1}{\mathrm{vol} (B_{R})}  
    \int_{B_{R}}\! e^{2\pi i k^{\star} y} \, \mathrm{d}y 
     \, =\, \frac{3\bigl(\sin(z) - z
       \cos (z)\bigr)}{z^{3}}
\]
with $z=2\pi\lvert{k_{}}^{\star}\rvert R$.
Figure~\ref{fig:primicodiff} shows sections through the corresponding
three-dimensional diffraction patterns, orthogonal to the fivefold,
threefold and twofold symmetry axes.\medskip

\begin{figure}[t]
\centering
\includegraphics[width=\columnwidth]{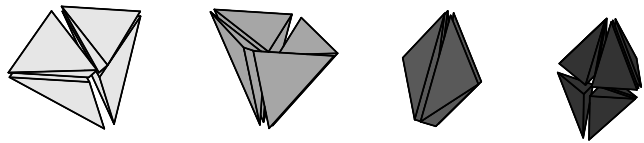}
  \caption{~The tiles of the Danzer tiling appear in groups of four 
           (A,B,C) or eight (K), forming (topological) octahedra.}
  \label{fig:abckoct}
\end{figure}

An example of an F-type (face-centred) icosahedral model set is
\emph{Danzer's tiling},\cite{Dan89} which was first constructed from
an inflation rule, and is also known as the \emph{ABCK tiling}, after
the labels Danzer used for the four tetrahedral prototiles. In the
ABCK tiling, the tetrahedral tiles always occur in the configurations
shown in Figure~\ref{fig:abckoct}, so one can alternatively work with
assembled prototiles consisting of four tiles of type A, B or C, and
eight tiles of type K.

\begin{figure}[t]
\centering
\includegraphics[width=\columnwidth]{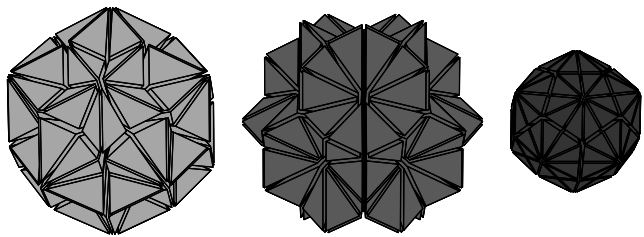}
\caption{~The three icosahedrally symmetric vertex stars of the Danzer
  tiling, comprising exclusively tiles of type B, C or K.}
  \label{fig:abckballs}
\end{figure}

The ABCK tiling is mutually locally derivable\cite{DPT93,Roth} from
the Socolar-Steinhardt tiling,\cite{SS86} so both describe equivalent
structures. An interesting property of Danzer's ABCK tiling is the
fact that it possesses particularly simple perfect local rules, which
can be formulated as purely geometric packing rules on the level of
the octahedra.\cite{Dan89} The Danzer tiling has three icosahedrally
symmetric vertex stars, each comprising just one type of tiles, which
are shown in Figure~\ref{fig:abckballs}.  Under inflation, these act
as seeds of globally icosahedrally symmetric Danzer tilings.

For the Danzer tiling comprising these larger prototiles of
Figure~\ref{fig:abckoct}, all vertices are located on the face-centred
icosahedral module
\[
   \mathcal{M}^{}_{\mathrm{F}} \, = \,
   \bigl\langle v^{}_{1}+v^{}_{2}, v^{}_{2} +
    v^{}_{3},v^{}_{3}+v^{}_{4},
    v^{}_{4} +v^{}_{5},v^{}_{5}+v^{}_{6},v^{}_{6}-v^{}_{1}  
    \bigr\rangle^{}_{\mathbb{Z}}\, ,
\]
which is a submodule of $\mathcal{M}^{}_{\mathrm{P}}$ of index
$2$. Explicitly, one has
\[
   \mathcal{M}^{}_{\mathrm{P}} \, = \,
   \mathcal{M}^{}_{\mathrm{F}} \cup 
    (\mathcal{M}^{}_{\mathrm{F}} + \tau^{2}u)\, ,
\]
where
$u=\frac{1}{2}(v^{}_{1}-v^{}_{2}+v^{}_{3}-v^{}_{4}+v^{}_{5}-v^{}_{6})
= (1,1,1)$. For this choice of coordinates, $u$ is \emph{not} in
$\mathcal{M}^{}_{\mathrm{P}}$.  The vertex point set can be described
as a three-component model set\cite{Roth,KPSZ94} based on a cut and
project scheme \eqref{eq:cps} with physical and internal space
$\mathbb{R}^3$. The corresponding lattice $\mathcal{L}$ is the
embedding of $\mathcal{M}^{}_{\mathrm{F}}$ in $\mathbb{R}^{6}$, which
is similar to the root lattice $D_{6}$. The vertices of the four types
of (topological) octahedra (thus disregarding their centres) separate
into three different types, which stem from different cosets of the
embedding lattice.

\begin{figure}[t]
\centering
\includegraphics[width=\columnwidth]{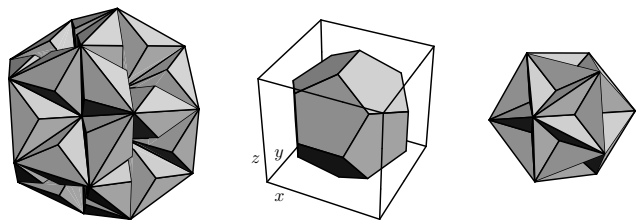}
  \caption{~Windows for the vertices of type I, II and III of the Danzer 
  tiling. They are shown in the correct relative size and orientation.}
  \label{fig:danwin}
\end{figure}

In fact, the usual description as a three-component model set uses the
projections of so-called `holes' in the lattice $\mathcal{L}$. Holes
are vertices of the Voronoi cells whose distance from points of the
lattice is a local maximum.\cite{CS} If the distance is an absolute
maximum, the hole is called deep, otherwise shallow. The vertices of
the Danzer tiling then fall into three groups: Vertices of type I are
projections from deep holes which lie in the coset
$\mathcal{L}+(1,1,1,\tau,\tau,\tau)$, those of type II from deep holes
in the coset $\mathcal{L}+(\tau,\tau,\tau,-1,-1,-1)$ and vertices of
type III from shallow holes in the coset
$\mathcal{L}+(\tau,0,1,-1,0,\tau)$. The corresponding three windows
have icosahedral symmetry and are shown in
Figure~\ref{fig:danwin}. The window for vertex type I is a
dodecahedral extension of an icosahedron, with pentagonal edge length
$2$ and volume $20(4-\tau)$, the window for vertex type II is a
dodecahedron of edge length $2/\tau$ and volume $4(\tau+2)$, and the
third window is a great dodecahedron (a Kepler-Poinsot polyhedron),
with pentagonal edge length $2$ and volume $20(\tau-1)$.  The
$\star$-map is the same as for the primitive model set above.

The diffraction pattern of the Danzer tiling has spots on the 
corresponding dual module
\begin{equation}\label{eq:danmod}
    \mathcal{M}^{\circledast}_{\mathrm{F}} \, = \, 
    \frac{1}{2(\tau+2)} \, \bigl( \mathcal{M}^{}_{\mathrm{P}} \, \cup \, 
    (\mathcal{M}^{}_{\mathrm{P}} + u)\bigr)
\end{equation}
with $u$ as above. Whereas the primitive tilings has diffraction spots
on $\mathcal{M}^{\circledast}_{\mathrm{P}}=\frac{1}{2(\tau+2)}
\mathcal{M}^{}_{\mathrm{P}}$ only, the Danzer tiling has additional
spots on the shifted copy $\frac{1}{2(\tau+2)}
(\mathcal{M}^{}_{\mathrm{P}} + u)$. Note that the union
$\mathcal{M}^{}_{\mathrm{P}} \cup (\mathcal{M}^{}_{\mathrm{P}} +
u)=\mathcal{M}^{}_{\mathrm{B}}$ corresponds to the body-centred
icosahedral module.

\begin{figure}[t]
\centering
\includegraphics[width=0.8\columnwidth]{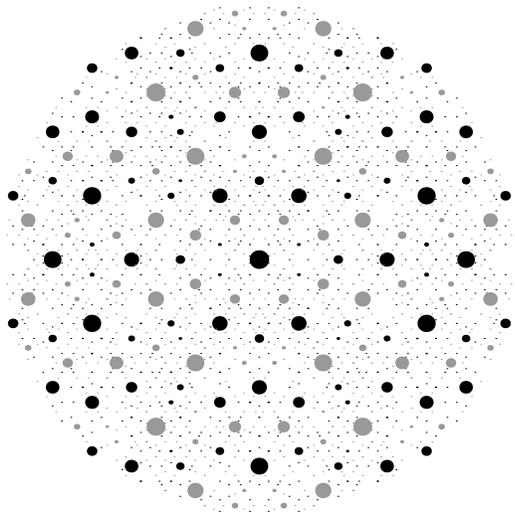}
  \caption{~Twofold section
   of the diffraction pattern of the ABCK tiling, with
   scatterers on all vertices of type II.}
  \label{fig:dantwodiff}
\end{figure}

Due to the relation between the symmetry directions and the shift $u$,
not all high-symmetry sections through the origin will show peaks from
both modules in Eq.~\eqref{eq:danmod}.  In fact, only the twofold
sections through the origin contain peaks from both parts in
Eq.~\eqref{eq:danmod} and thus display the full Fourier module, while
the three- and fivefold sections only contain peaks from
$\frac{1}{2(\tau+2)}\mathcal{M}^{}_{\mathrm{P}}$.
Figure~\ref{fig:dantwodiff} shows the twofold section of the
diffraction for a Dirac comb of vertex type II only, so the window is
simply a dodecahedron, which we approximate by a sphere of radius
$R=\bigl(\frac{3(\tau+2)}{\pi}\bigr)^{1/3}\approx 1.5118$. In
Figure~\ref{fig:dantwodiff}, the `black' diffraction spots belong to
$\frac{1}{2(\tau+2)}\mathcal{M}^{}_{\mathrm{P}}$, while the grey spots
belong to the coset. 

\begin{figure}[t]
\centering
\includegraphics[width=0.8\columnwidth]{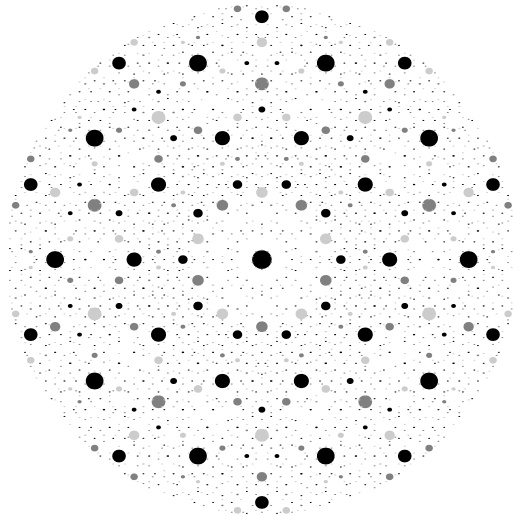}
  \caption{~Fivefold section
   of the diffraction pattern of the ABCK tiling; see text
   for details.}
  \label{fig:danfivediff}
\end{figure}

To visualise the diffraction along the fivefold axis, we combine the
section through the origin with two parallel sections containing the
coset reflections. The result is shown in
Figure~\ref{fig:danfivediff}. The spots in
$\frac{1}{2(\tau+2)}\mathcal{M}^{}_{\mathrm{P}}$ are again shown in
black, while the two different grey colours distinguish the spots in
the two parallel planes containing $u$ (dark grey) or $-u$ (light
grey). This pattern demonstrates that the overall rotational symmetry
here is fivefold (not tenfold) and inversion symmetric. The latter
property accounts for the tenfold rotation symmetry of the section
through the origin (black spots). Sections with threefold symmetry
display the analogous phenomena.

The distinction between the diffraction of a primitive and of a
face-centred icosahedral model set is thus immediately recognisable
from the spot locations in a twofold section. For further (practical)
details and examples we refer to the recent
literature.\cite{SDBook}\medskip

Within the realm of regular model sets, diffraction is thus pretty
well understood. We know that regular model sets are pure point
diffractive,\cite{Hof,Martin,Crelle} and Eqs.~\eqref{eq:modeldiff} and
\eqref{eq:modelamp} provide explicit expressions for the intensities
in terms of the Fourier transform of the window. Homometry of model
sets (within the same cut and project scheme) can be traced back to
equality of the covariogram of the window, and explicit examples of
homometric model sets have been constructed.\cite{BG07} Thermal
fluctuations can be taken into account in a fashion that is analogous
to the crystallographic case;\cite{Hof2,BBM} see
Section~\ref{sec:randis} below.

The Bragg diffraction has some robustness property beyond the class of
regular model sets. Recently, Strungaru\cite{S05} proved that, for any
Meyer set $\varLambda\subset \mathbb{R}^{d}$, the corresponding Dirac
comb $\omega = \delta^{}_{\varLambda}$ always shows a non-trivial
point diffraction, though in general the spectrum will be mixed and
not pure point.  However, the point part is substantial in the sense
that for any $\varepsilon > 0$, the set of peaks $\bigl\{ k \in
\mathbb{R}^{d} \mid \widehat{\gamma}\, (\{ k \}) \ge (1-\varepsilon)
\, \widehat{\gamma}\, (\{0\}) \bigr\}$ (all peaks with intensity near
the maximum intensity) is relatively dense. While we do not have a
complete answer to the question what structures are pure point
diffractive,\cite{BT} it is clear that a pure point spectrum imposes
strong constraints on the possible structures.\cite{BLR}

For the remainder of this article, we are looking at systems that show
continuous diffraction, both singular and absolutely continuous. The
discussion of examples with and without random disorder will shed some
light on the much more complex situation beyond the pure point
diffractive regime.

\section{Systems with continuous diffraction}
\label{sec:cont}

It seems a relatively recent experimental observation that diffuse
scattering (as an indication of structural disorder, and not just of
thermal fluctuations) is a widespread phenomenon.\cite{WW,Withers} It
is thus natural to also investigate continuous diffraction spectra
from a more mathematical perspective. Again, we briefly present
illustrative examples, most of which have been analysed completely and
rigorously by now.

\subsection{Singular continuous diffraction}

Let us begin by recalling the paradigm of singular continuous
diffraction, the \emph{Thue-Morse} (or Pruhet-Thue-Morse)
system.\cite{AS} It is usually defined via the substitution rule
$a\mapsto ab$, $b\mapsto ba$.  A bi-infinite fixed point sequence $w$
emerges from iterating the square of this rule with the legal seed
$a|a$. Define the Dirac comb
\[
    \omega \, = \, \sum_{n\in\mathbb{Z}} f(w(n))\, \delta_{n}\, ,
\]
where $f(a)=1$ and $f(b)=-1$. One can now show that the autocorrelation
measure exists\cite{Wie27,Mah27,K,BG08} and is of the form
\[
   \gamma \, = \, \sum_{m\in\mathbb{Z}} \eta(m)\, \delta_{m}\, ,
\]
with $\eta(0)=1$ and the recursion
\[
   \eta(2m)\, =\, \eta(m) \quad\text{and}\quad
   \eta(2m+1)\, = \, -\frac{1}{2}\,\bigl(\eta(m)+\eta(m+1)\bigr) ,
\]
which is valid for all $m\in\mathbb{Z}$. This exact
renormalisation-type structure is the golden key to prove the spectral
type \emph{and} to calculate the measure explicitly. 

\begin{figure}[t]
\centering
\includegraphics[width=0.8\columnwidth]{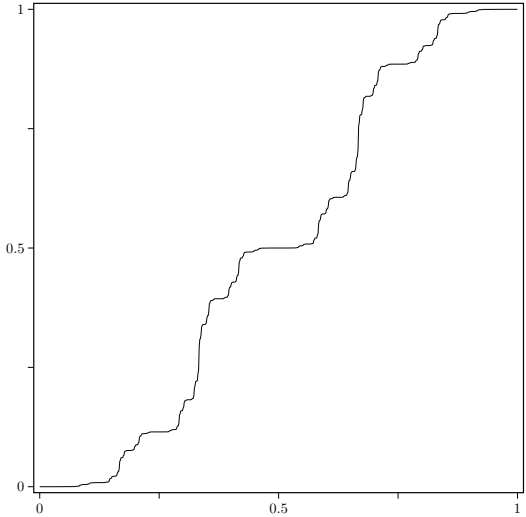}
  \caption{~Distribution function of the Thue-Morse diffraction measure
   on the unit interval.}
  \label{fig:tmmeasure}
\end{figure}

The diffraction measure is $1$-periodic,\cite{B02,BGG12} and hence of
the form $\widehat{\gamma}=\mu \ast \delta^{}_{\mathbb{Z}}$ with a
positive, singular continuous measure $\mu$. To describe the latter
explicitly, one defines the distribution function $F(x)=\mu([0,x])$ on
the unit interval. It is consistently extended to a function on
$\mathbb{R}$ by setting $F(x+n)=F(x)+n$ for $n\in\mathbb{Z}$. This
way, $F(x)-x$ is $1$-periodic and possesses the uniformly converging
Fourier series
\[
   F(x) - x \, = \, 
   \sum_{m=1}^{\infty}  \frac{\eta(m)}{m\pi}\, \sin(2\pi mx)\, .
\]
For computational purposes, however, it is advantageous to use an
approximation in terms of a uniformly converging sequence of
distribution functions as follows. Define $F^{}_{0}(x)=x$ and 
the functional iteration
\[
   F^{}_{N+1}(x) \, = \, \frac{1}{2}\int_{0}^{2x} 
   \bigl(1-\cos(\pi y)\bigr)\, \mathrm{d}F^{}_{N}(y)
\] 
for $N\ge 0$. Since this iteration maps distribution functions for
absolutely continuous measures to distribution functions of the same
type, one can write $\mathrm{d}F^{}_{N}(x)=f^{}_{N}(x)\, \mathrm{d}x$
with a Radon-Nikodym density $f^{}_{N}$. One can now check explicitly
that this leads to
\[
   f^{}_{N}(x) \, = \, \prod_{\ell=1}^{N} 
   \bigl(1-\cos(2^{\ell}\pi x)\bigr),
\]
where the empty product is to be evaluated as $1$. Since the densities
$f^{}_{N}$ become increasingly spiky (and do not converge as a
sequence of functions), one uses the distribution functions $F^{}_{N}$
to illustrate the resulting measure. Note that the sequence
$(F^{}_{N})^{}_{N\in\mathbb{N}}$ converges uniformly,\cite{BGG12} but
not absolutely. This is in line with the fact that $\mu$ is singular
continuous, and thus cannot be approximated by a norm-converging
sequence of absolutely continuous measures.\cite{RS} The resulting
limit distribution function $F$ is illustrated in
Figure~\ref{fig:tmmeasure}. Despite its similarity with the Cantor
measure of Figure~\ref{fig:cantor}, $F$ is a strictly increasing
function. This means that there is no proper plateau here.\medskip

\begin{figure}[t]
\centering
\includegraphics[width=\columnwidth]{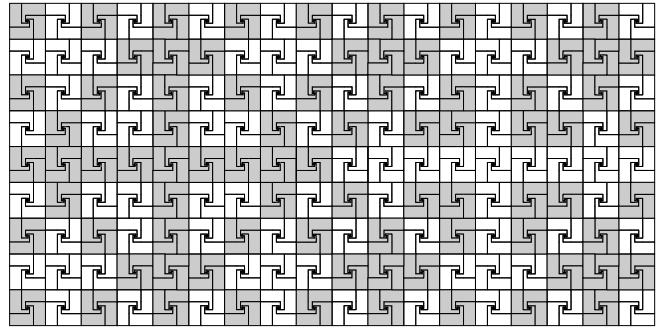}
  \caption{~Patch of the squiral tiling, obtained by two inflation steps 
   from the central seed, which is legal.}
  \label{fig:squiral}
\end{figure}

A non-trivial planar example emerges from the \emph{squiral} inflation
rule from Figure~10.1.4 in Gr\"{u}nbaum and Shephard.\cite{GS} It
effectively leads to an aperiodic $2$-colouring of the square lattice,
according to the chirality of the square dissections; see
Figure~\ref{fig:squiral} for an illustration. Positioning a point
measure of weight $1$ or $-1$ in the centre of the two types of
squares, one obtains a weighted Dirac comb with average weight
$0$. Due to the inflation structure, one can derive a recursion formula
for the corresponding autocorrelation.\cite{squiral,TAO} 

\begin{figure}[t]
\centering
\includegraphics[width=\columnwidth]{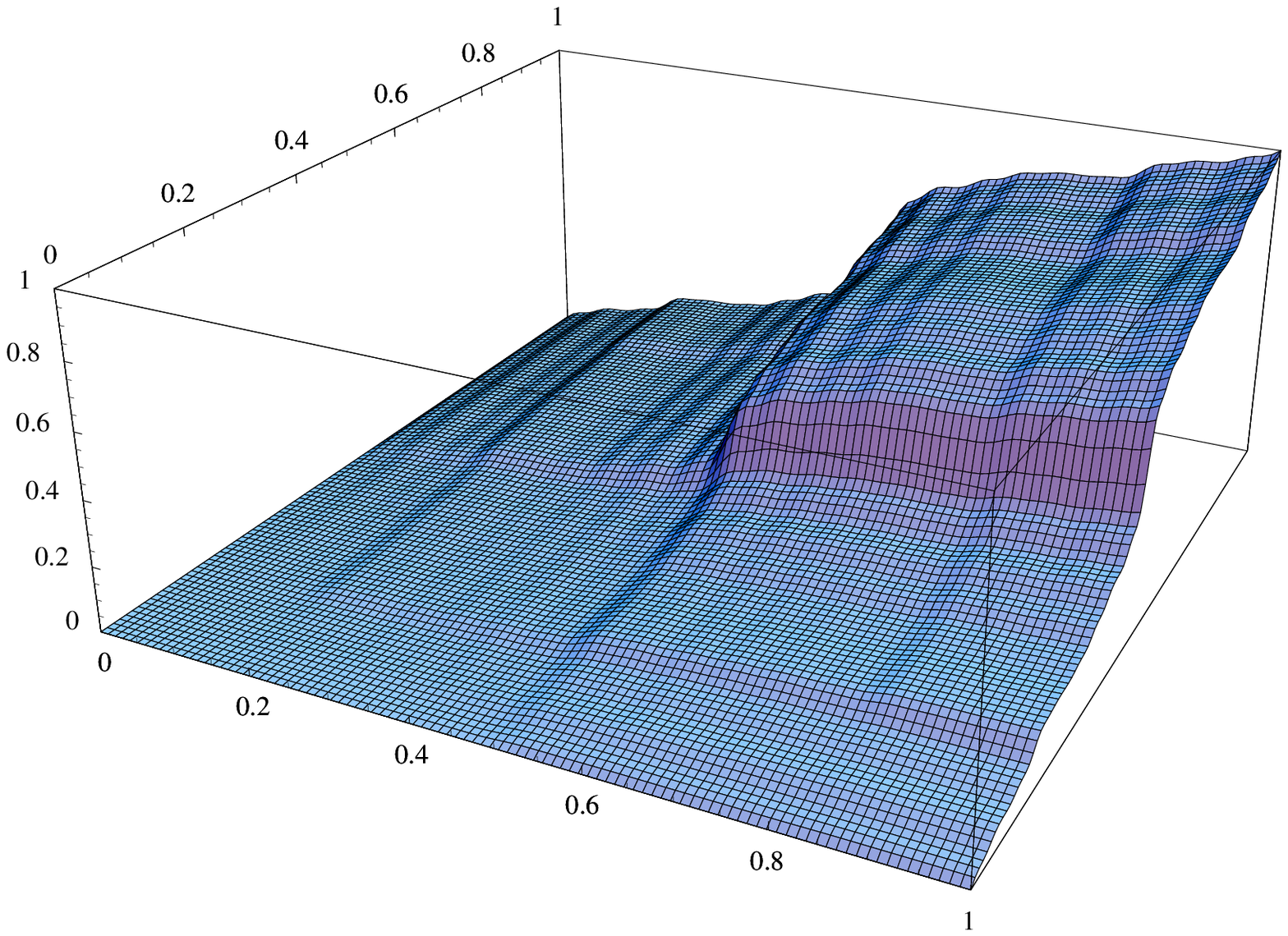}
\includegraphics[width=\columnwidth]{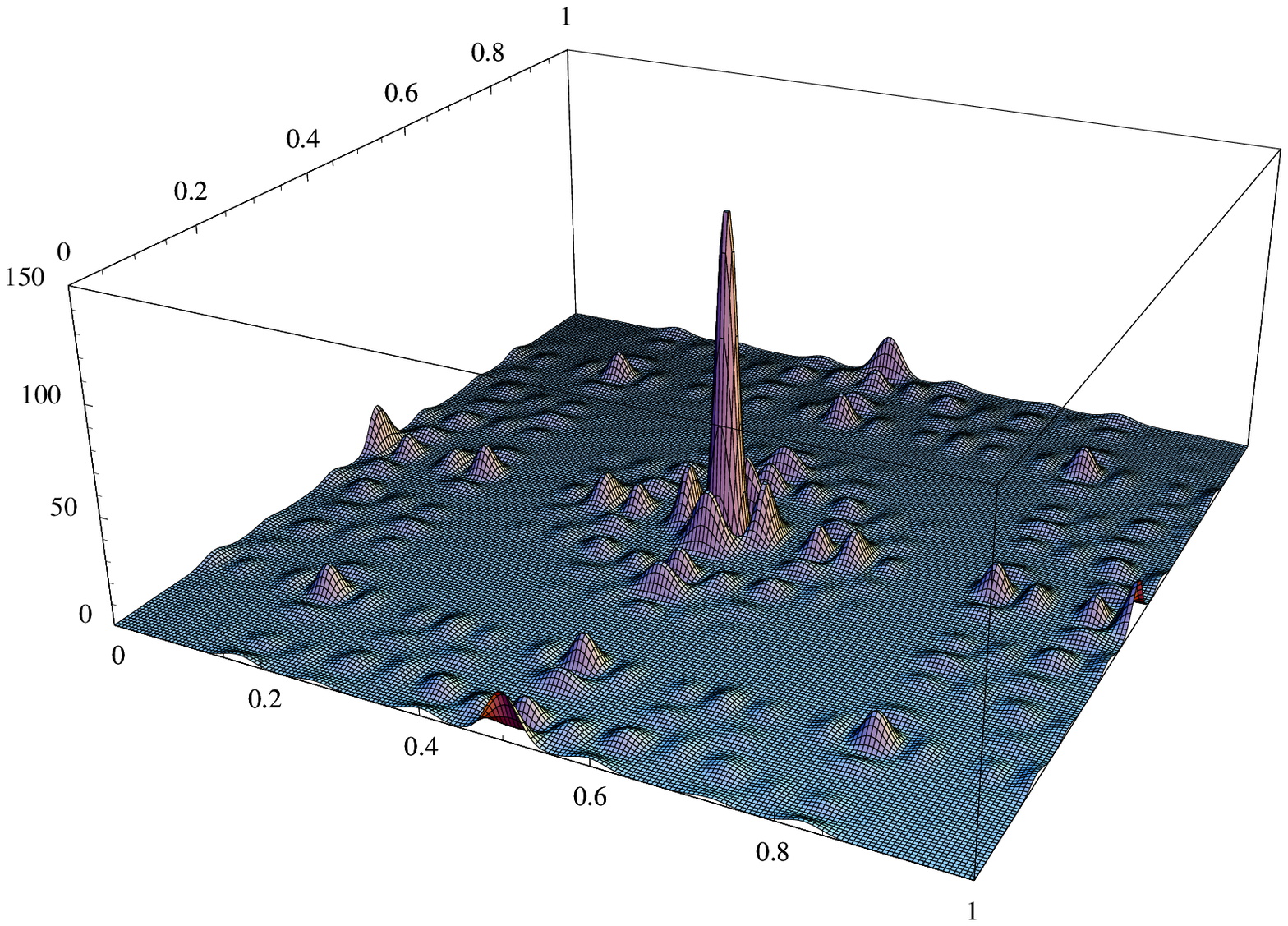}
\caption{~Third step of the Riesz product structure of
  Eq.~\eqref{eq:riesz} for the diffraction measure of the squiral
  tiling; see text for details.} \label{fig:squiraldiff}
\end{figure}

By constructive methods, in complete analogy to the case of the
Thue-Morse sequence, one can show that this Dirac comb leads to a
purely singular continuous diffraction measure.\cite{squiral,TAO} As
in the one-dimensional case, it can explicitly be calculated, and
represented as a two-dimensional Riesz product. The result reads
\begin{equation}\label{eq:riesz}
   f^{}_{N}(x,y) \, = \, \prod_{\ell=0}^{N-1} 
   \vartheta(3^{\ell}x,3^{\ell}y)\, ,
\end{equation}
where the function $\vartheta$ is defined by
\[
\begin{split}
   \vartheta (x,y) \, = \,  \frac{1}{9} \Bigl(&1
   + 2 \cos (2\pi x) + 2 \cos (2\pi y)\\
  & - 2 \cos \bigl(2\pi (x+y)\bigr) 
   - 2 \cos \bigl(2\pi (x-y)\bigr)\Bigr)^{2}\, .
\end{split}
\]
As in the one-dimensional case, the corresponding distribution
function possesses a uniformly convergent Fourier series
representation, which involves the autocorrelation coefficients. The
density function $f^{}_{3}$ (bottom) and the corresponding
distribution function $F^{}_{3}$ (top, normalised such that
$F^{}_{3}(0,0)=0$) are shown in Figure~\ref{fig:squiraldiff}.

\subsection{Rudin-Shapiro chain and its Bernoullisation}

A simple, deterministic system with continuous diffraction is the
\emph{binary Rudin-Shapiro chain}. While it is usually presented via a
four-letter substitution rule, the corresponding weighted Dirac comb
$\omega^{}_{\mathrm{RS}}=\sum_{n\in\mathbb{Z}} w(n)\delta^{}_{n}$ can
be defined by the sequence of weights $(w(n))^{}_{n\in\mathbb{Z}}$
with $w(n)\in\{\pm 1\}$, initial conditions $w(-1)=-1$, $w(0)=1$, and
the recursion
\begin{equation}\label{eq:rs}
   w(4n+\ell)=
    \begin{cases} w(n),  & \mbox{for $\,\ell\in\{0,1\}$,} \\
          (-1)^{n+\ell}\,w(n), & \mbox{for $\,\ell\in\{2,3\}$.}
     \end{cases}
\end{equation}
The arrangement of the two weights looks as follows
\[
  \includegraphics[width=\columnwidth]{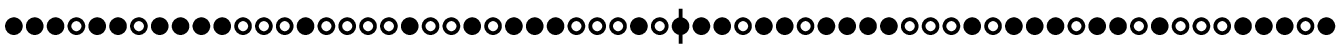}
\]
where the line denotes the origin, and filled (open) dots correspond
to weights $1$ ($-1$).

Despite the deterministic structure, the autocorrelation measure of the
balanced Dirac comb $\omega^{}_{\mathrm{RS}}$ (which has average
scattering strength $0$) can be shown\cite{Rudin,Shapiro,Q} to be
$\gamma^{}_{\mathrm{RS}} = \delta^{}_{0}$. A simple proof of this is
obtained by considering the induced recursion relation for the
autocorrelation coefficients.\cite{BG09,diffrev} The corresponding
diffraction measure is thus $\widehat{\gamma^{}_{\mathrm{RS}}} =
\lambda$, which is purely absolutely continuous, and shows no trace
whatsoever of the underlying deterministic order in the sequence. In
fact, the system is (almost surely, meaning for almost all
realisations of the random sequence) homometric with the random Dirac
comb on $\mathbb{Z}$ with weights from $\{\pm 1\}$ chosen at random,
independently at each position and with equal probability.

We can combine the deterministic sequence and independently chosen 
random numbers by considering the \emph{Bernoullisation} of the
Dirac comb $\omega^{}_{\mathrm{RS}}$, which we define as\cite{BG09}
\begin{equation}\label{eq:randrs}
   \omega_{p} \, = \sum_{n\in\mathbb{Z}} w(n)\, X(n)\,\delta_{n}\, .
\end{equation}
Here, $(w(n))^{}_{n\in\mathbb{Z}}$ is the binary Rudin-Shapiro
sequence of weights from Eq.~\eqref{eq:rs}, whereas
$(X(n))^{}_{n\in\mathbb{Z}}$ is an i.i.d.\ family of random numbers,
each taking values $1$ and $-1$ with probabilities $p$ and $1-p$ (so
$0\le p\le 1$), respectively. The limit cases $p\in\{0,1\}$ bring us
back to the deterministic Dirac comb $\pm\omega^{}_{\mathrm{RS}}$,
while the case $p=\frac{1}{2}$ corresponds to the Bernoulli comb with
weights $1$ and $-1$ mentioned above. The Bernoullisation thus
interpolates between the deterministic Rudin-Shapiro sequence and the
completely uncorrelated sequence of independent random numbers. It can
also be interpreted as a `model of second thoughts', where the sign of
the weight at position $n$ is changed with probability $1-p$.

Using the strong law of large numbers,\cite{Ete} it can be
shown\cite{BG09,TAO} that the autocorrelation $\gamma_{p}$ of the
Dirac comb $\omega_{p}$ is almost surely given by
\[
   \gamma_{p}  \, = \, (2p-1)^{2}\,\gamma^{}_{\mathrm{RS}} + 
   4 p (1-p)\, \delta^{}_{0}
   \, = \, \delta^{}_{0}\, ,
\]
\emph{irrespective} of the value of the parameter $p\in [0,1]$. So the
diffraction of this Dirac comb, for any choice of the parameter $p$,
is (almost surely) $\widehat{\gamma_{p}}=\lambda$, and the entire
family of Dirac combs is homometric. 

This simple example highlights the fact that diffraction in general
cannot distinguish `order' in the sense of a deterministic structure
from that in the presence of entropy. Note that the deterministic
Rudin-Shapiro sequence has zero entropy, while the Bernoulli comb has
entropy $\log(2)$, which is the maximum entropy for a binary
sequence. For general $p$, the entropy is $H(p) = - p\log (p) -
(1\!-\!p) \log (1\!-\!p)$, so it varies continuously between $0$ and
$\log(2)$. Regardless, the diffraction of all these combs is the
same. This result provides a glimpse at how degenerate, and hence
difficult, the inverse problem can be in the presence of continuous
spectra. Similar arguments can be used in higher dimensions (in
particular by considering product structures), and examples in two
dimensions involving lower rank entropy have also been
discussed.\cite{BW10,BG12}

\subsection{Random displacements and thermal fluctuations}
\label{sec:randis}

There are various important applications of Bernoulli-type disorder in
real systems. The most obvious one is known as the \emph{random
  occupation model}, which covers lattice gases and models of chemical
disorder. Traditionally, this has been formulated for lattice-based
systems only,\cite{C,Welberry} but the corresponding results hold in
much greater generality. This includes model sets,\cite{BM98} but also
structures with a substantial degree of positional
disorder.\cite{K1,K2,BBM} It turns out that the lattice assumption can
be replaced by rather general principles from probability theory that
revolve around the strong law of large numbers.\cite{Ete}

This change of perspective is also of value for the treatment of the
effects of thermal fluctuations to the diffraction of solids. In fact,
rather than restricting to a lattice and small vibrations in a
harmonic potential, the famous Debye-Waller contribution\cite{C} can
alternatively be derived from the assumption that the scatterers are
randomly displaced from their equilibrium positions, independently of
each other, but based on the same probability distribution. This opens
the door to another application of the strong law of large numbers, as
was first observed by Hof.\cite{Hof2} Two further advantages are the
validity for considerably more general point sets than lattices and
the independence of the argument of the small displacement
assumption. At least for sufficiently high temperatures, this
alternative approach is reasonable.

Consider a Delone set $\varLambda\subset\mathbb{R}^{d}$ that is
sufficiently nice (where we refer to the literature\cite{Hof2,BBM} for
the precise conditions). In particular, we assume that the Dirac comb
$\delta^{}_{\varLambda}$ possesses the autocorrelation $\gamma$. The
random displacement is described as 
\[
   \varLambda^{\!\prime} \, = \, \{ x+t_{x}\mid x\in\varLambda \}\, ,
\]
where $(t_{x})^{}_{x\in\varLambda}$ is a family of i.i.d.\ random
translation vectors with common probability distribution $\nu$.
Then, with probability one, $\delta^{}_{\varLambda^{\!\prime}}$ has the autocorrelation
\begin{equation}\label{eq:gamprime}
   \gamma^{\;\prime} \, = \, \gamma \ast (\nu \ast \widetilde{\nu}) 
   \, + \,\mathrm{dens}(\varLambda) \, (\delta^{}_{0} - \nu \ast \widetilde{\nu})\, .
\end{equation}
The corresponding diffraction is obtained by Fourier transform and reads
\begin{equation}\label{eq:gamprimdiff}
   \widehat{\gamma^{\;\prime}} \, = \, \lvert\widehat{\nu}\rvert^{2} \,
   \widehat{\gamma} \, +\,  \mathrm{dens}(\varLambda) 
   \bigl( 1 - \lvert\widehat{\nu}\rvert^{2} \bigr).
\end{equation}
Here, $\widehat{\nu}$ is a uniformly continuous function on
$\mathbb{R}^{d}$ that vanishes at infinity, and the formula holds
almost surely, as Eq.~\eqref{eq:gamprime}. If $\widehat{\gamma}$, the
diffraction of $\delta^{}_{\varLambda}$, is a pure point measure, the
pure point part of $\widehat{\gamma^{\;\prime}}$ is given by
$\lvert\widehat{\nu}\rvert^{2}\, \widehat{\gamma}$ (hence by a
modulation of the intensities, which is the Debye-Waller factor),
while the continuous part is $\mathrm{dens}(\varLambda) (\delta^{}_{0}
- \nu \ast \widetilde{\nu})$.  Note, however, that
Eq.~\eqref{eq:gamprimdiff} is by no means restricted to pure point
diffractive systems. An explicit dependence on the temperature can be
modelled by the appropriate choice of the displacement distribution
$\nu$. Further details and generalisations are discussed in the
literature.\cite{K1,BBM}

\subsection{Random tilings}

Random tilings form a particularly interesting and relevant class of
structures, as was early pointed out by Elser.\cite{Els85} The
structure of the various ensembles and their diffraction is not as
well understood as in the deterministic case, though a fairly complete
picture was sketched by Henley.\cite{Henley} From a physical point of
view, most results are `clear', on the basis of convincing (scaling)
arguments from statistical mechanics.  The mathematical counterpart,
however, is still incomplete, and various properties have escaped a
proof so far, particularly in dimensions $2$ and higher. In fact, it
is a characteristic feature of random tilings to show a strong
dependence on the dimension, as we will illustrate by some examples.
\medskip

\begin{figure}[t]
\centering
\includegraphics[width=0.8\columnwidth]{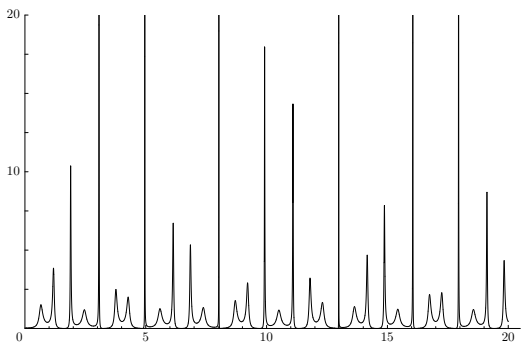}
\caption{~Continuous part of the diffraction pattern of a Fibonacci
  random tiling. The range for the wave number $k$ on the horizontal
  axis is the same as in Figure~\ref{fig:fibodiff}.}
  \label{fig:fibort}
\end{figure}

Let us first consider a random version of the Fibonacci chain.  Here,
one starts with two prototiles as before (one interval of length
$\tau$ and one of length $1$), and builds a tiling of $\mathbb{R}$ by
choosing them with probabilities $p$ and $1-p$, where $p=\tau^{-1}$
leads (almost surely) to realisations with the same relative tile
frequencies as the deterministic chain of
Figure~\ref{fig:fibocps}. Due to the linear arrangement, the ensemble
is well under control by elementary methods from probability
theory. In particular, one can either invoke the ergodicity of the
Bernoulli (coin tossing) chain\cite{BH} or the renewal
theorem\cite{BBM} to show that the random Dirac comb obtained this way
almost surely leads to the diffraction measure
\begin{equation}\label{eq:fibort-diff}
    \widehat{\gamma} \; = \, 
     \biggl(\frac{\tau + 2}{5}\biggr)^{\! 2} \delta^{}_{0}
    \, +\,  h(k) \, \lambda
\end{equation}
with the Radon-Nikodym density function
\[
  h(k) \, = \, \frac{\tau+2}{5}\,\frac{(\sin (\pi k/\tau))^{2} }
      {\tau^{2} \, (\sin(\pi k \tau))^{2} + \tau \, 
            (\sin (\pi k))^{2} - (\sin (\pi k/ \tau))^{2} }\, .
\]
The factor $(\tau+2)/5\approx 0.7236$ is the density of the
corresponding point set, which equals that of the deterministic
counterpart discussed earlier. Apart from the trivial Bragg peak at
$k=0$, the diffraction is thus absolutely
continuous. Figure~\ref{fig:fibort} shows the function $h$, which is
smooth but still shows a spiky structure that resembles the pure point
diffraction of the perfectly ordered Fibonacci chain from
Figure~\ref{fig:fibodiff} to an amazing degree.

The mechanism behind the absolutely continuous nature of the
diffraction in Eq.~\eqref{eq:fibort-diff} can be understood as
follows. Due to the choice of the intervals, each realisation can be
lifted within the cut and project scheme of the perfect Fibonacci
chain of Figure~\ref{fig:fibocps}. Almost surely, one then obtains a
sequence of lattice points that deviate from the perfect case via
fluctuations that diverge linearly with the system 
size.\cite{Henley,HoeffeDiss}
This destroys the coherence needed for Bragg peaks (at $k\ne 0$) or
singular continuous contributions to $\widehat{\gamma}$. \medskip

\begin{figure}[t]
\centering
\includegraphics[width=0.8\columnwidth]{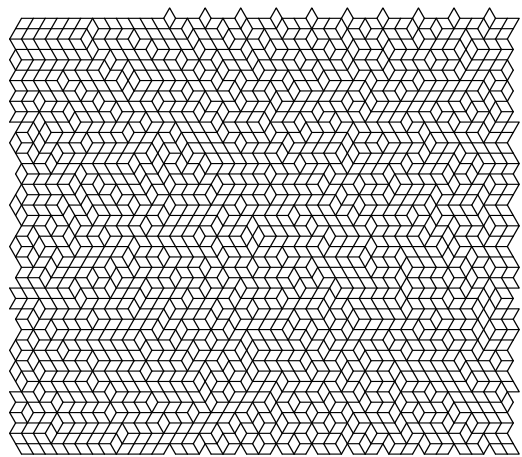}
  \caption{~Typical patch of a rhombus (or lozenge) random tiling,
    with periodic boundary conditions. Here, the vertical rhombus 
    is less frequent than the other two types, hence breaking the 
    statistical threefold symmetry.}
  \label{fig:hexpatch}
\end{figure}

Random tilings in the plane show a different behaviour, which also
depends on the symmetry. In particular, it is important whether one
deals with a crystallographic symmetry (such as statistical three- or
sixfold symmetry in the lozenge tiling) or not (such as statistical
eightfold symmetry in the random octagonal tiling). An example of the
former case, with broken symmetry, is illustrated in
Figure~\ref{fig:hexpatch}.  The underlying ensemble is well studied in
statistical physics.\cite{Kas63,DG,Ken97,HoeffeDiss,Ken00}

The lozenge (or rhombus) with opening angle $\pi/3$ occurs in three
possible orientations in all typical lozenge random tilings (which are
subject to the condition that any resulting tiling is face to face and
covers the plane without overlaps). One can now use the relative
frequencies of the three prototiles to parametrise the ensemble. By
purely group theoretic methods, one can then show that the entropy has
a unique maximum at the (unique) point of maximal
symmetry.\cite{Ric99,Henley} This shows an interesting entropic
mechanism for the stabilisation of tilings with statistical
symmetry. The value of the entropy (calculated per tile) is known
exactly from a mapping to the two-dimensional antiferromagnetic Ising
model on the triangular lattice, which was exactly solved by
Wannier.\cite{Wan}

The underlying ensemble is special also in the sense that one does not
only know the free energy and the entropy, but also the two-point
correlation functions, at least asymptotically.  Since this is the
autocorrelation of the system, when placing point scatterers of unit
mass on each vertex point, the diffraction measure for almost all
realisations of the lozenge random tiling (with edge length $1$, say)
is of mixed type, and has the form $ \widehat{\gamma} =
\bigl(\widehat{\gamma}\bigr)_{\mathrm{pp}} + \bigl(
\widehat{\gamma}\bigr)_{\mathrm{ac}}$. The pure point part is\cite{BH}
\begin{equation}\label{eq:hexfou}
    \bigl(\widehat{\gamma}\bigr)_{\mathrm{pp}} \; = \; 
    \frac{4}{3}\sum_{\scriptscriptstyle (k_1,k_2)\in \varGamma^*}
   \bigl((-1)^{k_1}\rho_1^{}+(-1)^{k_2}\rho_2^{}+\rho_3^{}\bigr)^2 
   \delta_{(k_1,k_2)}^{} \, ,
\end{equation}
where $\varGamma^{*}$ is the dual lattice of the triangular lattice,
spanned by $v^{}_{1} = \bigl(1,-\frac{1}{\sqrt{3}}\bigr)$ and
$v^{}_{2} = \bigl(0,\frac{2}{\sqrt{3}}\bigr)$, and
$(k^{}_{1}, k^{}_{2})$ is a shorthand for the wave vector $k^{}_{1}
v^{}_{1} + k^{}_{2} v^{}_{2} \in \varGamma^{*}$.  The pure point part
reflects the underlying lattice structure,\cite{B02} while the
absolutely continuous one is the fingerprint of the structural
disorder. It is effectively repulsive in nature, as expected, which
manifests itself\cite{BH,BS} in the property that the diffuse
intensity is `repelled' by the Bragg peaks.

\begin{figure}[t]
\centering
\includegraphics[width=0.65\columnwidth]{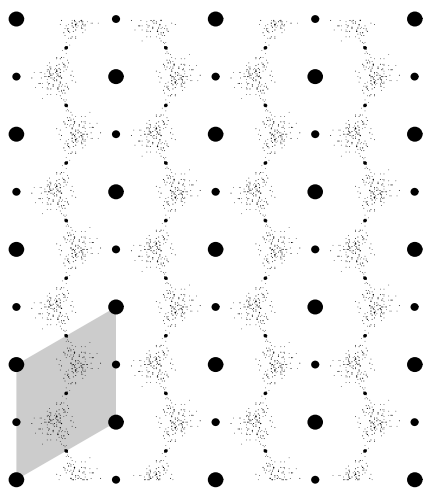}
  \caption{~Diffraction pattern of the lozenge random tiling of
  Figure~\ref{fig:hexpatch}. The pattern is lattice periodic,
  with the shaded rhombus as a fundamental domain.}
  \label{fig:hexfou}
\end{figure}

\begin{figure}[b]
\centering
\includegraphics[width=0.6\columnwidth]{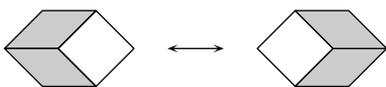}
  \caption{~A simpleton flip used in the thermalisation of the
            Ammann-Beenker (or octagonal) tiling.}
  \label{fig:flip}
\end{figure}

The diffraction of the example from Figure~\ref{fig:hexpatch} is shown
in Figure~\ref{fig:hexfou}. The pattern is lattice periodic.  The pure
point part (big spots) follows from the exact formula in
\eqref{eq:hexfou}, while the continuous part (small spots) was
calculated numerically by H\"{o}ffe\cite{BH,HoeffeDiss} via FFT
techniques.\medskip

\begin{figure}[t]
\centering
\includegraphics[width=0.75\columnwidth]{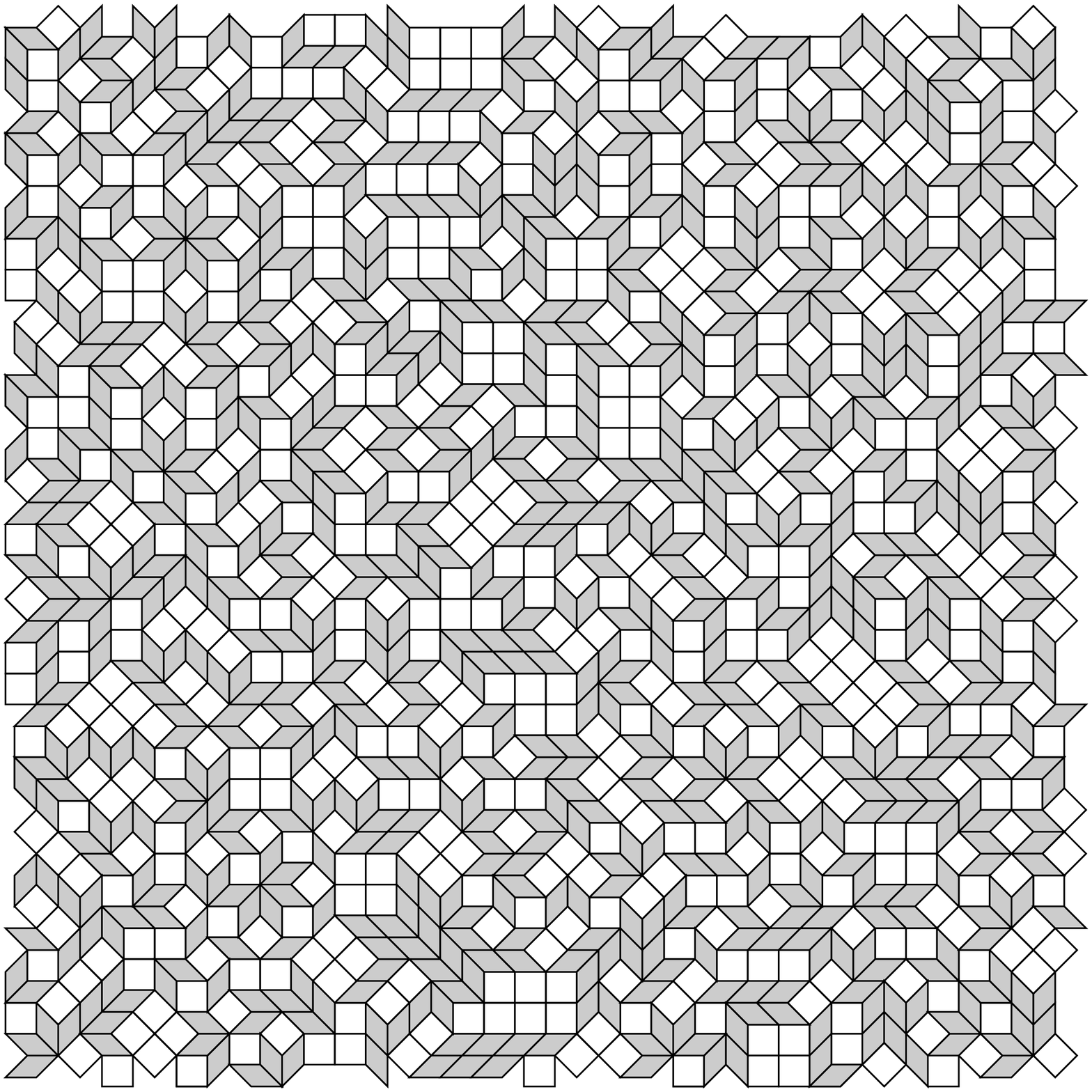}
  \caption{~Patch of an octagonal random tiling, obtained
  by thermalisation of a periodic approximant via simpleton flips.}
  \label{fig:abrandom}
\end{figure}

The corresponding situation for the randomised Ammann-Beenker or
octagonal tiling looks similar at first sight, and leads (via
simpleton flip thermalisation, see Figure~\ref{fig:flip}) to patches
of the form shown in Figure~\ref{fig:abrandom}. However, the possible
vertex positions are no longer restricted to a lattice, but only to
the module $\mathbb{Z} [\xi^{}_{8}]$ with $\xi^{}_{8}$ a primitive
$8$th root of unity. This module is the corresponding set of
cyclotomic integers and a dense point set in the plane, as explained
earlier. As a result, apart from the trivial Bragg peak at $0$, the
diffraction measure will be continuous, with singular and absolutely
continuous components. The reason behind this is the logarithmically
diverging fluctuation of the embedding surface from the deterministic
surface of the model set relative.\cite{Henley} Due to the larger
positional freedom of the vertex points, this fluctuation is strong
enough to destroy the coherence that is needed for non-trivial Bragg
peaks, but not strong enough to avoid singular continuous
contributions.\cite{HoeffeDiss}

\begin{figure}[t]
\centering
\includegraphics[width=0.8\columnwidth]{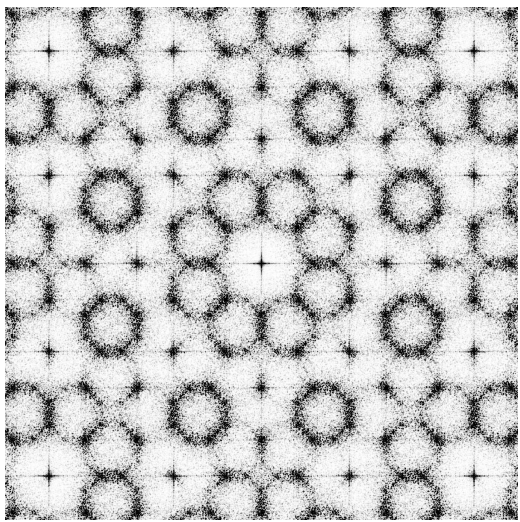}
\caption{~Numerical approximation to the diffraction image of the
  random tiling of Figure~\ref{fig:abrandom}.}
  \label{fig:abranddiff}
\end{figure}

Unfortunately, this is one of the claims that have not yet been
proved, though there can be hardly any doubt about its correctness.  A
numerical calculation\cite{TAO} of the diffraction of the finite patch
shown in Figure~\ref{fig:abrandom} leads to the pattern of
Figure~\ref{fig:abranddiff}, a similar result was obtained by
H\"{o}ffe\cite{HoeffeDiss} via FFT. A comparison with the diffraction
of the perfect Ammann-Beenker tiling in Figure~\ref{fig:abdiff} still
reveals a lot of similarities, despite the approximative nature of the
calculation. In particular, one can clearly map the strong peaks of
the perfect case to positions of the random tiling diffraction, and
also various ring-type structures are clearly common to both images.
In view of these similarities, it is not clear to what extent kinematic
diffraction of a \emph{finite} patch can distinguish the perfect from
a random tiling.

The simpleton flip of Figure~\ref{fig:flip} provides a standard
approach for the preparation of random tiling samples. It works well
also for other tilings with rhombic prototiles, where one might have
different types of simpletons to consider (for instance, there are two
such configurations in the rhombic Penrose tiling).  One usually
starts from a periodic approximant (to minimise boundary effects) to a
perfect tiling, which is not difficult to construct, and runs the
simpleton flip thermalisation until correlations have decayed. In such
ensembles, the process can be shown to be topologically transitive, so
that the entire ensemble compatible with these boundary conditions is
accessible.\cite{Henley,HoeffeDiss} Note, however, that there are
other important ensembles, such as the random square triangle tilings,
where no such local flip exists. Here, one needs alternative methods,
such as the well-studied `zipper' move\cite{OH} that temporarily
introduces some new (auxiliary) tiles that enable a randomisation
path, until the created tiles annihilate themselves again and leave a
modified square triangle tiling behind.  \medskip

Finally, the case of random tilings in $3$-space is clearly of great
interest. A natural candidate from the very
beginning\cite{Els85,Henley} has been the randomised version of the
primitive icosahedral tiling, which is built from the two rhombohedra
of Figure~\ref{fig:keprhomb}.  While there are $24$ complete vertex
configurations in the Kramer-Neri projection tiling, counted up to
icosahedral isometries, there are $5450$ possible ones in a typical
random tiling.\cite{BBKS93} So, it is clear that this version locally
shows a much higher degree of disorder.  However, unlike the previous
examples, the fluctuations away from the embedding hypersurface seem
to be bounded,\cite{Els85,Henley} which implies a diffraction of mixed
type, this time with a pure point and an absolutely continuous
component --- despite the statistical icosahedral symmetry, which is
non-crystallographic; a numerical confirmation was obtained by
Monte-Carlo simulation techniques.\cite{Tang90}

If one employs a statistical variant of the projection method, the
fluctuations mentioned above lead to a distribution in internal space
that can be described by a density function. The latter will resemble
a Gaussian profile,\cite{Els85,Henley} which makes the pure point part
of the diffraction explicitly accessible via an appropriate extension
of the model set theorem to this case.\cite{Richard} This gives
diffraction formulas of PSF type where the sums on both sides run over
dense point sets.  A further generalisation was recently formulated
for measures by Lenz and Richard.\cite{LR}

\section{Outlook}

The discovery of quasicrystals\cite{SBGC84} in 1982 had a profound
impact on various disciplines, including mathematics and, in
particular, to harmonic analysis and mathematical diffraction
theory. The approach described above emerged from the investigation of
aperiodically ordered systems, and offers a method that can be applied
to a wide range of structures.

After 30 years of quasicrystal research, the diffraction of
mathematical quasicrystals that are described by cut and project sets
(regular model sets) is well understood. Such structures are pure
point diffractive, much as conventional crystals, except that the
Bragg peaks are supported on a point set that is dense in space. For
many standard examples, the corresponding diffraction amplitudes can
be calculated explicitly, for instance in terms of Fourier transforms
of the corresponding window(s).

The situation changes quickly if one leaves the realm of model sets.
As discussed, Meyer sets still inherit some of the structure, in the
sense that their diffraction measure contains non-trivial pure point
components. For substitution (or inflation) based structures, examples
with all spectral types are known. In this article, we met examples of
all three types --- the Fibonacci chain (which is a pure point
diffractive model set), the Thue-Morse chain (which has singular
continuous diffraction) and the Rudin-Shapiro chain (with absolutely
continuous spectrum). In fact, it is easy to come up with a
substitution system that has a mixed spectrum comprising all three
spectral types.

Quasicrystals are expected to contain some inherent (or structural)
disorder, and it is therefore desirable to understand the effect of
disorder on diffraction, and, vice versa, the conclusions on disorder
that one can draw from examining diffraction patterns, in particular
with regard to continuous diffraction. This is far from being well
understood, but the examples discussed above provide a glimpse at the
general situation. As the Bernoullisation example shows, diffraction
cannot always detect the nature of `order', for instance whether the
latter is of deterministic or entropic origin.  Conversely, diffuse
diffraction does not always need to be a sign of random disorder. At
present, we only have a very limited knowledge of how large the
homometry classes can be. In the pure point diffractive case, a recent
approach by Lenz and Moody\cite{LM,LM2} provides one possibility for
an abstract parametrisation. Unfortunately, this approach does not
seem to be extendable to cover continuous diffraction components.  The
investigation of further examples with different types or degrees of
order will hopefully shed more light on this matter.

\begin{figure}[t]
\centering
\includegraphics[width=0.7\columnwidth]{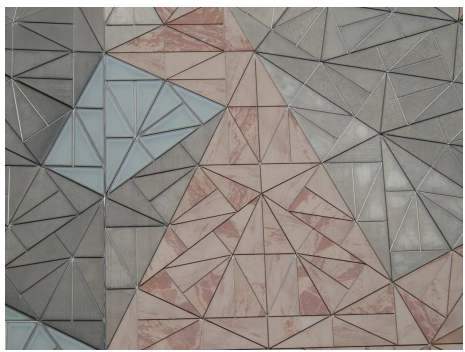}
\caption{~Detail of a fa\c{c}ade at Melbourne's Federation Square
featuring a pinwheel tiling. Photography \copyright\ U.~Grimm.}
  \label{fig:pinwheel}
\end{figure}

One does not have to go far to find examples of important, yet still
not completely understood systems. A prominent one is the Conway-Radin
pinwheel tiling.\cite{Rad94} This tiling is based on a single
triangular prototile (of edge lengths $1$, $2$ and $\sqrt{5}$), with
an inflation rule of linear inflation multiplier $\sqrt{5}$, so each
re-scaled triangle (which is planar) is dissected into five congruent
copies. Figure~\ref{fig:pinwheel} shows a photograph of a patch of the
tiling, which has been used as a theme for Melbourne's Federation
Square development. Because the inflation contains a rotation that is
incommensurate with $\pi$, a new direction is introduced in each
inflation step. Consequently, each infinite pinwheel tiling contains
triangles in infinitely many distinct orientations, and the
corresponding tiling space even has complete circular
symmetry.\cite{Rad94,Rad97,MPS06} The diffraction patterns shows striking
similarity to a powder diffraction from a square-lattice based
structure.\cite{BFG07a} While there is strong evidence for
sharp rings in the diffraction pattern (which are singularly
continuous in the plane), mimicking the case of the rotation-averaged
square-lattice structure, the presence of further rings or absolutely
continuous components is still unclear.\medskip

More generally, one needs a unified setting for the diffraction of
systems with mixed spectra. An interesting suggestion was made by
Gou\'{e}r\'{e}\cite{Gou} on the basis of the intensity measure of the
Palm measure of a point process. This provides an alternative way to
define the autocorrelation of the system. It is possible to include
cases such as crystallographic systems or model sets into this
scheme,\cite{LM,LM2} and it was recently also shown\cite{BBM} how to
use this approach in a systematic way for systems with various kinds
of disorder. Since the theory of point processes is a highly developed
branch\cite{DVJ1,DVJ2} of modern probability theory, the use of these
methods looks rather promising.

\subsection*{Acknowledgements}

It is our pleasure to thank Franz G\"ahler and Peter Zeiner for
comments and discussions. This work was supported by the German
Research Council (DFG), within the CRC 701.

\footnotesize{

}
\end{document}